\documentclass[preprint,12pt,3p,authoryear]{elsarticle}

 \biboptions{comma,round}

\graphicspath{{figures/}{../figures/}}
\usepackage[space]{grffile}  % this enables spaces in graphics file names

\usepackage[T1]{fontenc}

\usepackage{mathptmx} % to get Times New Roman similar font
\usepackage{natbib} 
\usepackage{latexsym}
\usepackage{textcomp}
\usepackage{longtable}
\usepackage{multirow,booktabs}
\usepackage{amsfonts,amsmath,amssymb}
\usepackage[utf8]{inputenc}
\usepackage[english]{babel}
\addto\captionsenglish{}

\usepackage{soul}
\setstcolor{red}
\usepackage{siunitx}
\usepackage[version=4]{mhchem}
\usepackage{subfiles}
\usepackage{pdflscape}
\usepackage[labelfont=bf]{caption}

\usepackage[inline]{enumitem}
\newlist{mylist}{enumerate*}{1}
\setlist[mylist]{label=(\arabic*)}

\usepackage[section]{placeins}
\makeatletter
\AtBeginDocument{%
  \expandafter\renewcommand\expandafter\subsection\expandafter{%
    \expandafter\@fb@secFB\subsection
  }%
}
\makeatother

\usepackage{hyperref}
\hypersetup{colorlinks=true,allcolors={blue},pdfborder={0 0 0}}

\journal{Icarus}

\begin{document}

\begin{frontmatter}

%% Title, f and addresses


\title{Planet Four: A Neural Network's Search For Polar Spring-time Fans On Mars}

\cortext[cor1]{Corresponding author}
\author[unisa]{Mark D. McDonnell}
%\ead{mark.mcdonnell@unisa.edu.au}
\address[unisa]{Computational Learning Systems Laboratory, UniSA STEM, University of South Australia, Mawson Lakes, SA 5095, Australia}

\author[unisa]{Eriita Jones}
\author[qub]{Megan E. Schwamb\corref{cor1}}
\ead{m.schwamb@qub.ac.uk}
\address[qub]{Astrophysics Research Centre, School of Mathematics and Physics, Queen's University Belfast,Belfast BT7 1NN, UK}
\author[lasp,fub]{K-Michael Aye}
\address[lasp]{Laboratory for Atmospheric and Space Physics, University of Colorado at Boulder, Boulder, CO 80303, USA}
\address[fub]{Institute for Geological Sciences, Freie Universit\"at Berlin, Germany}
\author[lasp,dlr]{Ganna Portyankina}
\address[dlr]{Institute of Planetary Research, German Aerospace Center (DLR), Berlin, Germany.}
\author[psi]{Candice J. Hansen}
\address[psi]{Planetary Science Institute, 1700 E. Fort Lowell, Suite 106, Tucson, AZ 85719, USA}

\begin{abstract}
Dark deposits visible from orbit appear in the Martian south polar region during the springtime. These are thought to form from explosive jets of carbon dioxide gas breaking through the thawing seasonal ice cap, carrying dust and dirt which is then deposited onto the ice as dark `blotches', or blown by the surface winds into streaks or `fans'. We investigate machine learning (ML) methods for automatically identifying these seasonal features 
in High Resolution Imaging Science Experiment (HiRISE) satellite imagery. We designed deep Convolutional Neural Networks (CNNs) that were trained and tested using the catalog generated by Planet Four, an online citizen science project mapping the south polar seasonal deposits. We validated the CNNs by comparing their results with those of ISODATA (Iterative Self-Organizing Data Analysis Technique) clustering and as expected, the CNNs were significantly better at predicting the results found by Planet Four, in both the area of predicted seasonal deposits and in delineating their boundaries. We found neither the CNNs or ISODATA were suited to predicting the source point and directions of seasonal fans, which is a strength of the citizen science approach. The CNNs showed good agreement with Planet Four in cross-validation metrics and detected some seasonal deposits in the HiRISE images missed in the Planet Four catalog; the total area of seasonal deposits predicted by the CNNs was 27\% larger than that of the Planet Four catalog, but this aspect varied considerably on a per-image basis.
\end{abstract}%

\begin{keyword}
Mars \sep Machine learning \sep Classification \sep Mars, surface 


\end{keyword}
%\tableofcontents

\end{frontmatter}
 %\linenumbers
\section{Introduction}\label{S:1}

\begin{figure}
\centering
{\includegraphics[width=0.8\linewidth]{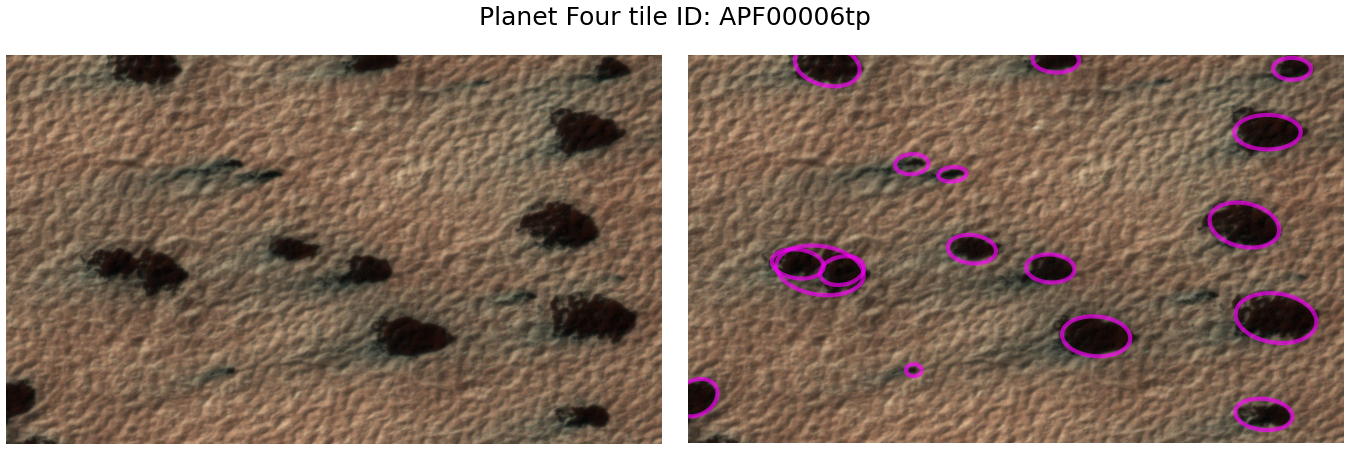}   \label{fig:00a} }
{\includegraphics[width=0.8\linewidth]{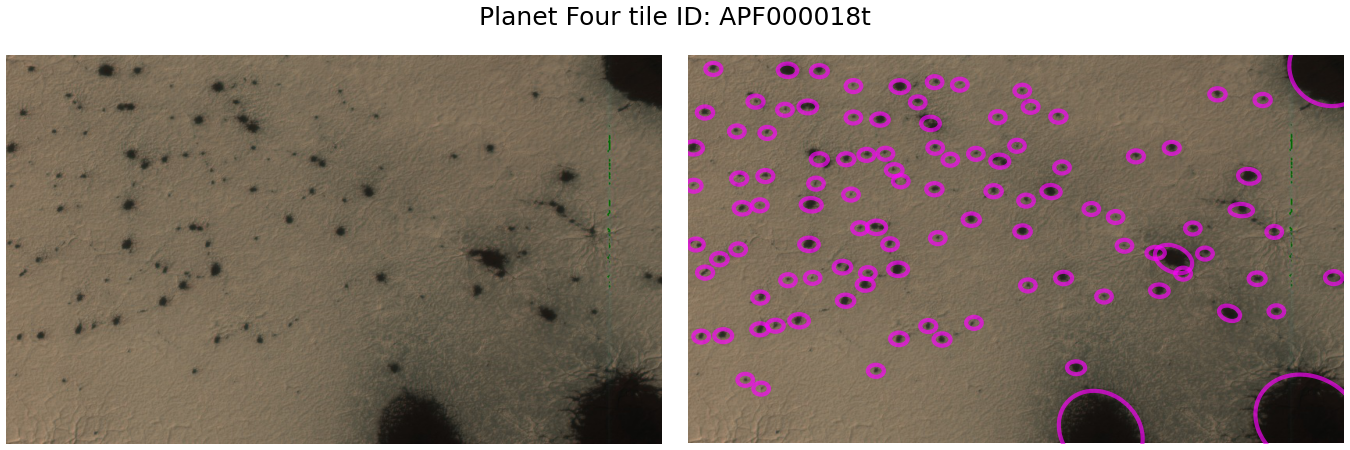}  
  \label{fig:00b}}
{ \includegraphics[width=0.8\linewidth]{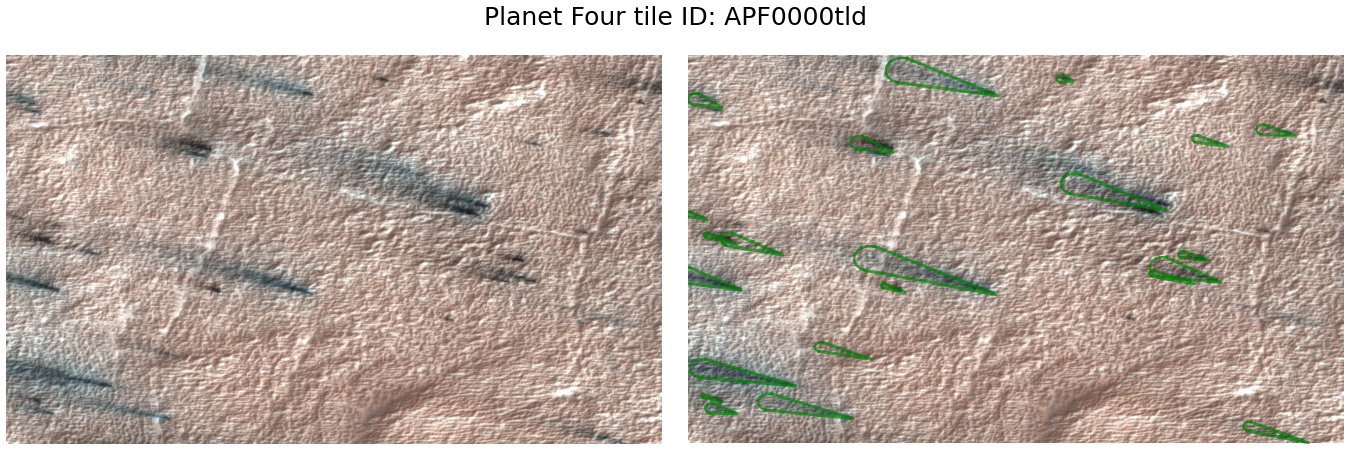}
  \label{fig:00c}}
 { \includegraphics[width=0.8\linewidth]{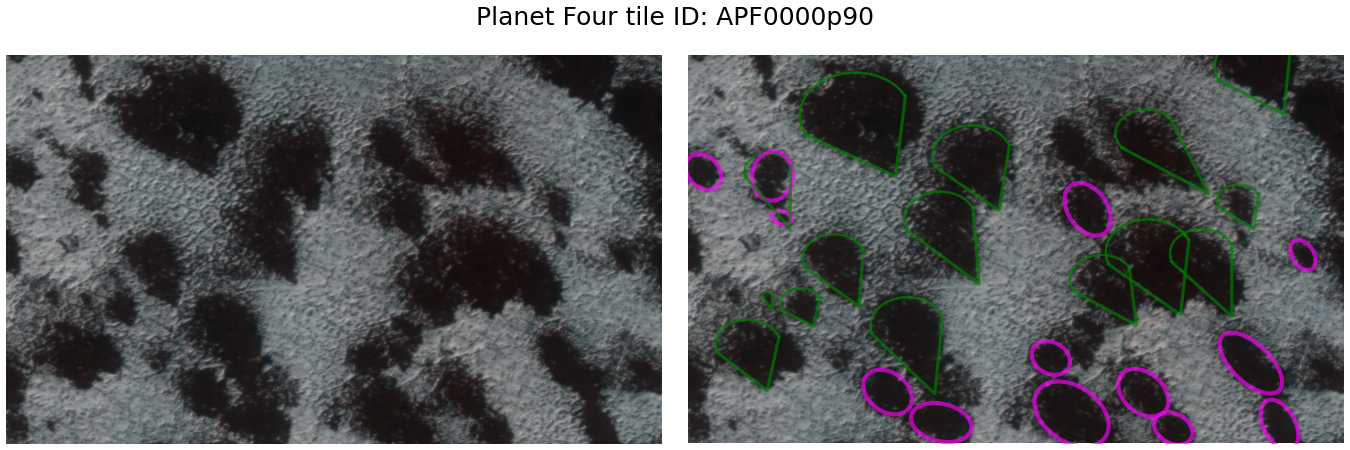}
  \label{fig:00d}}

\caption{Example tiles (HiRISE subimages) reviewed by the Planet Four project, on the left. On the right,  the resulting Planet Four fan (green) and blotch (magenta) catalog derived from the volunteer classifications are overlaid on the tile. Each tile is 648 pixels high and 840 pixels wide, but its ground resolution varies with HiRISE binning modes. These tiles are derived from HiRISE images \url{ESP_012008_0975} (P4 tile APF00006tp), \url{ESP_012889_0985} (APF000018t), \url{ESP_020780_0930} (APF0000tld), and \url{ESP_021491_0950} (APF0000p90). Note an example of overlap between a fan and two blotches in the lower right image, on its right hand side, above center. The primary purpose of this figure is to illustrate the two types of markings that humans could label for Planet Four, and the difference between them, namely fans and blotches. It should be noted that not all source data and volunteer markings are as well delineated as shown here---see, e.g.~Figures~\ref{fig:hirise_vs_p4_vs_ml} and ~\ref{fig:P4Miss}.} 
\label{fig:P4}
\end{figure}

Springtime on the Martian south polar region is marked by the appearance of dark streaks dotting the surface of the thawing carbon dioxide seasonal ice cap. The prevailing winds and explosive carbon dioxide (CO$_2$) gas jets that are breaking through the seasonal ice are thought to be jointly responsible for these surface features~\citep{2000mpse.conf...93K,piqeux2003,2006Natur.442..793K, 2007JGRE..112.8005K,2008JGRE..113.6005P,2010Icar..205..296T,2010Icar..205..311P,2013JGRE..118.2520P}.  In this currently favored model, first proposed by~\cite{2000mpse.conf...93K}, the jets transport dust and dirt from below the semi-translucent seasonal ice sheet up to the surface  where it is then distributed by the local surface winds and deposited back onto the ice as the dark seasonal fans visible from orbit (see Figure~\ref{fig:P4}). Laboratory experiments have been able to trigger dust eruptions from a layer of dust inside a carbon dioxide slab ice under Martian conditions, supporting this argument~\citep{Kaufmann2017118}.

Exploring the distribution of the seasonal fans provides valuable insights into surface wind patterns, the CO$_2$ jet process, and the climate cycles of Mars. This necessitates mapping the observable dark deposits, examining their recurrence over successive spring-times, and monitoring for the detection of newly emerged fans over an area of 1000s of square kilometers and hundreds of high-resolution satellite images.  The flotilla of spacecraft in orbit around Mars have captured the appearance and evolution of these springtime seasonal deposits. Hundreds to thousands or more seasonal fans are visible in high-resolution imagery taken during the southern spring~\citep{2003JGRE..108.5084P, 2010Icar..205..283H,2010Icar..205..296T,Aye19}. Identifying and mapping these seasonal fans is a difficult task, and no automated routine for doing so exists~\citep{Piqueux08,Aye19}.

To this end, the Planet Four\footnote{\url{http://www.planetfour.org}} citizen science project has crowd-sourced the identification and labeling of these spring-time seasonal deposits in over 200 High Resolution Imaging Science Experiment (HiRISE) \citep{McEwen07} camera images to create a catalog of $\sim$400,000 south polar seasonal fan deposits~\citep{Aye19} (see Figure~\ref{fig:P4}). Although a success, it can take a long time before sufficient numbers of annotations from volunteers are generated. A potential alternative or complement to crowd-sourced annotations that would speed up mapping considerably is the use of  machine learning.  In particular, tremendous advances have been made in recent years using supervised learning in the form of deep Convolutional Neural Networks (CNNs) applied to imagery. In supervised learning, a model is created by a training algorithm that learns to associate data samples with labels for those samples. Subsequently, the trained model is applied to the task of predicting labels for new data for which no pre-existing labels are available. Some examples of the use of deep CNNs in the planetary and space sciences includes crater detection/counting on the moon~\citep{Silburt19,Yang20} and Mars~\citep{Lee19}, detection of galaxies~\citep{Wu18,Walmsley19}, galaxy morphology classification~\citep{Walmsley19}, detection and classification of lunar rockfalls~\citep{Bickel19}, classification of terrain features on Mars~\citep{Wagstaff18,wagstaff2021mars}, and detection of changes that have occurred in images of the same surface location taken at two different times~\citep{Kerner.19}. A recent study   brought together citizen science and machine learning~\citep{Jones20}, as we also do in this paper.

For supervised learning methods to generalise effectively to new data, ideally both high volume and high diversity of data is available for training. However, a barrier to achieving this ideal is that human-labeling of large data sets can often be very costly to acquire.  The publicly available dataset represented by the \cite{Aye19} Planet Four catalog matches both of these conditions, while already having been labeled by humans. Therefore, we set out to design supervised machine learning methods that learn from the catalog, and hence enable automated detections of the presence and location of seasonal fan deposits in new Mars polar imagery from HiRISE.  As a step towards automating the mapping of seasonal fans, in this paper we present results from algorithms that aim to identify which specific pixels in images of the Mars surface belong to seasonal fans. To validate that our methods were well designed and that sophisticated methods such as CNN are warranted, we compare their predictions with those of the long-existing and still widely applied Iterative Self-Organizing Data Analysis Technique (ISODATA) clustering method~\citep{Ball} applied to the same target, i.e.~directly to the HiRISE imagery. We do not attempt to separate the predicted pixels into separated shapes nor to identify directionality. As would be expected given the success of deep CNNs on similar tasks, we found that the deep CNN we trained resulted in much better agreement with the Planet Four catalog than ISODATA clustering, and was capable of making predictions for some image portions missed by the Planet Four catalog.   

As a secondary study using supervised machine learning, we trained a model that aimed to determine if {\em any} CO$_2$ jet seasonal deposits  exist on the ``tiles'' (640$\times$824 pixel overlapping crops extracted  from full HiRISE images) that were inspected by human volunteers for Planet Four. This binary classification model achieved a high degree of accuracy, meaning that this approach has strong potential for narrowing down which tiles human volunteers are asked to look at, by omitting a large percentage of those predicted to show no CO$_2$ jet seasonal deposits, or at least prioritising the ones that do.

Our two types of models together suggest there is  strong potential for using machine learning to maximize the effort of the volunteers performing the human review of Mars imagery. Moreover, the fact that we did not identify an accurate way for machine learning to identify directionality of  CO$_2$ jet seasonal deposits  indicates that the collective effort of human volunteers remains as a vital resource.

The paper is organised as follows. In Sections~\ref{S:2a} and~\ref{S:2b} we outline the data that we used, and overviews our primary approach -- semantic segmentation -- and measures for assessing its accuracy. Next, in Section~\ref{S:ISODATA} we describe the baseline ISODATA method, and then Section~\ref{method-ml} presents our primary methods and results, i.e. use of supervised learning to train a deep CNN to semantically segment markings in HiRISE images. This firstly includes explanation of our design for each of the primary sub-tasks required by a supervised learning investigation, i.e.~how we split data into training and validation sets, how we trained an algorithm on the training set, and how inference is carried out using the trained model applied to validation data. The section then presents our results. Next, Section~\ref{method-ml-classify} describes methods and results for a secondary study  in which we train CNNs as binary classifiers that identify whether dark seasonal fans or blotches are present within a HiRISE image or subimage. This differs from the semantic segmentation approach in that it does not aim to identify the image pixels that belonged to fans and blotches.  Finally, Section~\ref{S:4} presents discussion of our results and conclusions.  Our python code for training and validating our models, and producing Figures 6--14 in this paper, is available on Zenodo: 
\href{https://doi.org/10.5281/zenodo.4292195}{doi:10.5281/zenodo.4292195}.

\section{Data}\label{S:2a}

\subsection{HiRISE Images}\label{S:HiRISE}

\begin{figure}
\centering
\includegraphics[width=0.42\columnwidth]{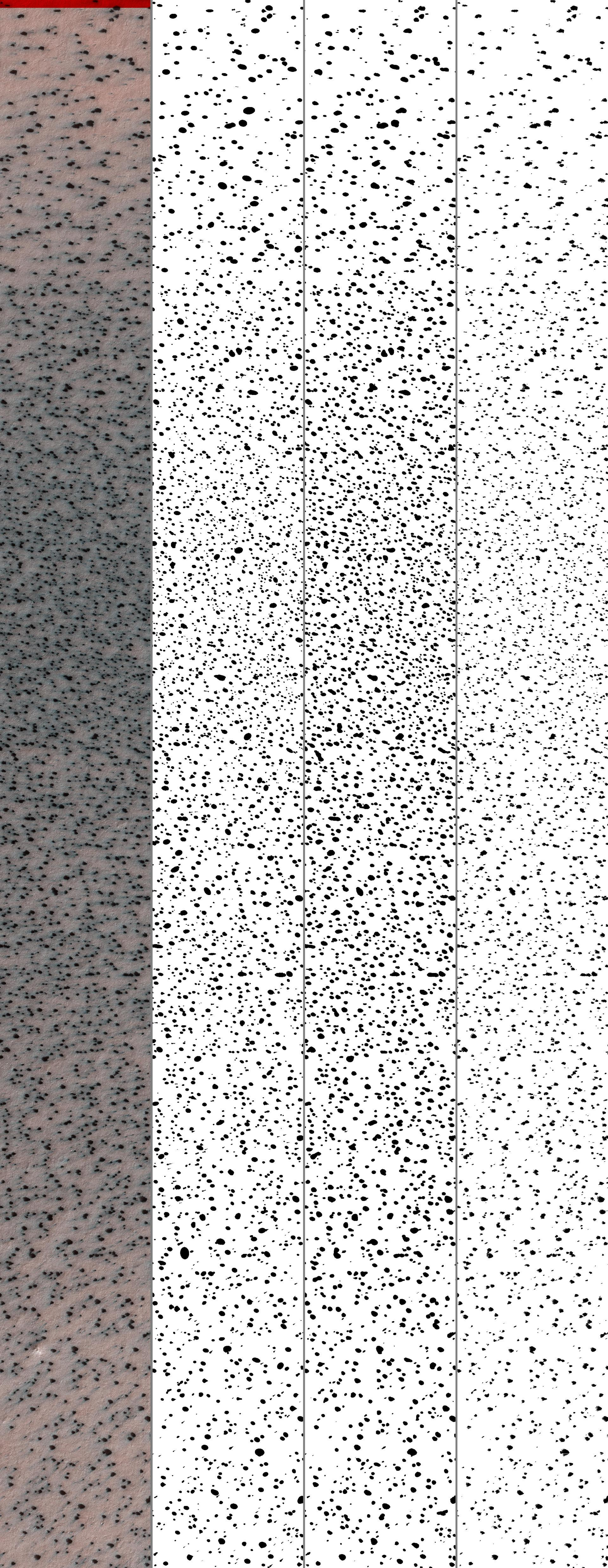}\quad\quad
\includegraphics[width=0.42\columnwidth]{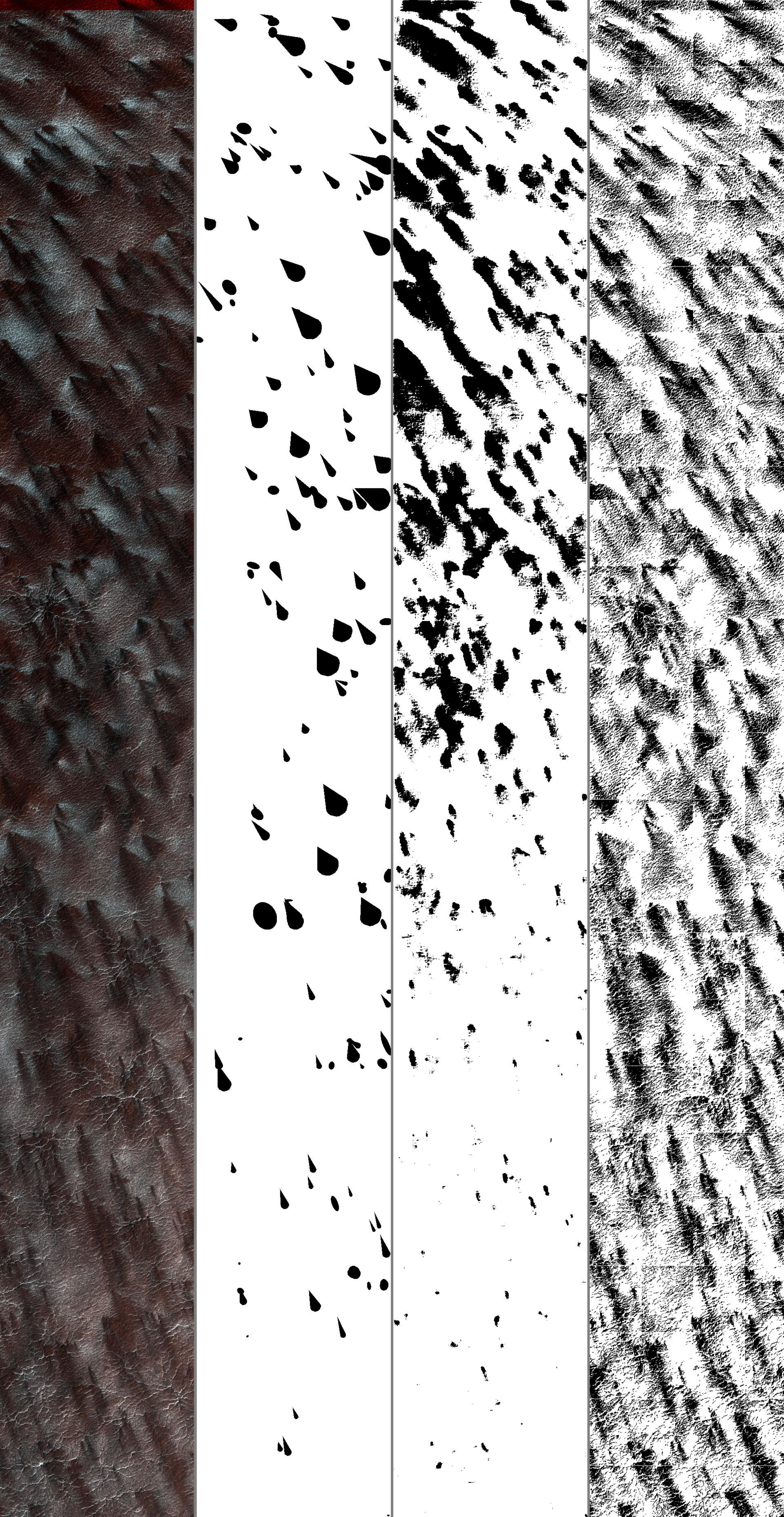}
\caption{Examples (left: HiRISE Image ID \url{ESP\_012008\_0975}, right: HiRISE Image ID \url{ESP\_020954\_0935}) from the set of  221 HiRISE images used. This figure qualitatively shows the nature of the full HiRISE images, the Planet Four catalog, and the variability of our methods' performance even within a single image.  Each example shows the original image  (left) with binary feature masks created from the P4 catalog, that we used for supervised learning (center-left), the binary result from our trained CNN's prediction in cross-validation (center-right), and the binary result from ISODATA clustering (right).}
\label{fig:feature_masks_on_HiRISE}
\end{figure}

The imagery used in this paper was sourced from 221 publicly  available\footnote{\url{https://www.uahirise.org/catalog/}} full color (RGB) images, acquired by the Mars Reconnaissance Orbiter's HiRISE \citep{2007JGRE..112.5S02M} camera in southern spring during Mars Years 29 and 30, the same as listed in Tables~1 and~2 in~\cite{Aye19}. These 221 images have pixel sampling scales of either 25, 50, or 100 cm per pixel. The size of the original 221 images were all either 1012 (pixel scale 100 cm), 2024 (pixel scale 50 cm) or 4048 pixels wide (pixel scale 25 cm). The heights vary in each image, ranging from a minimum of 10000 to a maximum of 80000 pixels. See Figure~\ref{fig:feature_masks_on_HiRISE} for two example full size  HiRISE images (left quarter of each example).

The 221 images can be grouped into subsets in various ways such that images in a group share similarities.  We chose to group by {\em region} as defined in Table~1 of~\cite{Aye19}; each region is indicative of a particular range of polar latitudes and longitudes that has been targeted for imaging multiple times by HiRISE. There are 28 distinct regions covering our data. The number of images in a region varied from 1 (5 instances) to 18.

\subsection{The Planet Four Catalog: Crowd-sourced Identification of Seasonal Fans}%
\label{sec:p4intro}
The Planet Four citizen science project has reviewed the set of HIRISE images that comprise our sample. It was decided that there was no need to compensate for directional bias, due to several reasons: 1.\ Due to natural orbit progression of the spacecraft, the angle of the imaging scan on the surface changes over time; 2.\ in most locations, local topography is dominating local wind direction, and is variable over the season, but to varying degree between regions of interest; 3.\ as the season progresses, the sun angles also change for the given local time at which most images are taken due to the spacecraft’s orbit. These variations randomize the final appearance of fan directions in image tiles of the Citizen Science display system.
Combining the multiple volunteer assessments together,~\citet{Aye19} has produced a catalog of carbon-dioxide jet produced markings in our sample of HiRISE images. We use this catalog of marking locations, shape, and sizes as our supervised learning labels for training and testing the ML deep CNNs, and we then compare various metrics of agreement between catalog and predictions made by the CNNs and ISODATA clustering. It should be noted that there are non-idealities in using the catalog as labels for supervised learning, which we detail in Section~\ref{S4truth}. The Planet Four data is previously published as downloadable Supporting Information~\citep{Aye19}\footnote{Also available online: \href{https://data.zooniverse.org/planet_four/P4_catalog_v1.1_L1C_cut_0.5_fan.csv.zip}{Link to Fan catalog}, \href {https://data.zooniverse.org/planet_four/P4_catalog_v1.1_L1C_cut_0.5_blotch.csv.zip}{Link to Blotch catalog}}. 

Planet Four volunteers reviewed ``tiles'', 648 pixels high and 840 pixels wide subframes that were cropped from the 221 HiRISE images, with adjacent tiles sharing 100 pixels of overlap, as detailed previously~\citep{Aye19}.  %JPEG conversion and cropping was performed using ImageMagick's ``convert'' command. 
In this tiling scheme, a tile had to fit entirely within the bounds of the original image. Consequently, a strip of pixels down the right hand side and at the base of the HiRISE image were not part of any tile and were never labeled by the Planet Four project. We removed these unlabeled areas of the original images by cropping, and thereby they did not contribute to any results reported in this paper. See Figure~\ref{fig:P4} (left) for example tiles cropped out of  full-size HiRISE, to this size of 648 $\times$ 840 pixels, as viewed by human labelers. For the sample of HiRISE images used in this work, 42,904 tiles were searched for seasonal fans by the Planet Four project. For further details, the reader is referred to \cite{Aye19}.

Visitors to the Planet Four website were tasked with identifying and marking any seasonal features present in the HiRISE subframes with drawing tools in the web interface. Human annotations were collected by Planet Four: {\em blotches} and {\em fans}---see Figure~\ref{fig:P4}. Blotches are ellipses represented with five parameters: a pixel height and width location for the center of the ellipse, a minor axis radius, a major axis radius, and an angle from horizontal. Fans are comprised from a semi-circle joined to an isosceles triangle. They are represented by an ice-cream-cone shape with five parameters: a pixel height and width location for the apex of the triangle, the vertex angle of the triangle, the angle between the horizontal and the line segment from the apex to the center of the triangle's base, and the distance from the apex to the center of the base. For a diagram of these parameters, see Figure 12 in~\citet{Aye19}. Volunteers were encouraged to draw with the fan tool, if clear directionality and a starting point is visible. Otherwise, Planet Four labelers were encouraged to use the blotch drawing tool. 

The human annotations collected by Planet Four \st{(based on typically  30 to 100 human reviews per image)} were combined together for each tile to identify the seasonal features present. A minimum of 30 up to 100 reviews per image tile were required before retiring an image tile. A clustering algorithm was employed to take the independent markings drawn by each volunteer and produce locations (mid-point for the blotches and starting point for fans) and outlines of the seasonal features based on the fan or blotch shape. A minimum of 3 markings within variable pixel distances were required for the density-based clustering scheme to have them entered into the final pool for averaging and catalog entry. Sources where 50\% or more of the volunteer-drawn markings were made with the fan tool were deemed to be fans, with the ice cream cone shape generated; otherwise an ellipse (a blotch is generated). Details of the clustering algorithm and validation of the resulting catalog are described in~\cite{Aye19}. For our sample of HiRISE images, the Planet Four project has produced a catalog of 159,558 fans and 250,164 blotches (ellipses), identifying locations of seasonal surface deposits produced by the CO$_2$ jet processes occurring during spring in the Martian south polar region~\citep{Aye19}. For this work, we utilize both the Planet Four fan and blotch catalogs, subsequently referring to these collectively as the Planet Four catalog.  Example fans and ellipses from the Planet Four catalog can be seen in the right hand column of Figure~\ref{fig:P4}.  CO$_2$ jet seasonal deposits  also can be seen in the center left strips in Figure~\ref{fig:feature_masks_on_HiRISE}, but the scale of that figure is not designed to clearly show individual fan/blotch shapes.

\subsection{Data Preprocessing}

The original 221 images are stored in RGB JPEG2000 format. Although this standard uses 16 bits per pixel per channel, the HiRISE image pixel values had a maximum value requiring only 10 bits per pixel per channel. For ease of use with training algorithms and to reduce RAM usage, we converted all images to 8 bits per pixel per channel by casting all pixel values to 32 bit floating point representations, multiplying by $255/2^{10}$ and then casting to 8-bit unsigned integers. While this changes the maximum value of the data by a factor of four, a linear rescaling will result in the same image dynamic range, but with four times fewer discrete pixel values. We found that this process does not typically affect the shape of the histograms of pixel values for each color when applied to HiRISE images. Moreover, the rescaling is justified by the fact that Planet Four also used HiRISE images converted to 8-bit per channel PNG images for human inspection.

\section{Separating \texorpdfstring{CO$_2$}{CO2} Jet Seasonal Deposits from Background}\label{S:2b}

\subsection{Semantic Segmentation}

As mentioned, recent work has used deep CNNs to identify craters on the moon~\citep{Silburt19,Yang20} and Mars~\citep{Lee19}. Our focus here is similar in the sense of aiming to automatically identify surface features. However, there is a crucial difference in the data available to us; the mentioned papers all benefit from elevation data, whereas our data source is high resolution color satellite imagery. In particular, the Mars Digital Terrain Model of~\cite{Lee19} has a resolution of 200m, and hence its pixel size is two orders of magnitude larger than our data. Even if the markings in the P4 catalog exhibited elevation differences, 200m pixel sizes would be  insufficient to resolve the majority of the catalog. As will be discussed, as well as the absence of elevation data, other non-idealities in the data results in a more challenging task for a deep CNN to learn than identification of craters based on elevation data.

There are different ways in which the Planet Four catalog\footnote{\url{https://www.zooniverse.org/projects/mschwamb/planet-four/about/results}} might be used for providing labels for supervised learning. The data in the catalog includes location of fans, their size, and their orientation. Considering that the overarching scientific questions relate to wind speed and direction, in our preliminary exploratory data analysis we attempted to train mask-RCNN models~\citep{MaskRCNN} to segment each individual instance of fans and masks, hoping that the results could be used to predict the extent of directionality and size of fans. However, the accuracy achieved was considered to be inadequate; validation predictions suggests the model was unable to learn the difference between fans that were highly directional, and  CO$_2$ jet seasonal deposits  of similar appearance with no directional features. The mask-RCNN models also struggled with cases where fans in the catalog overlapped (in some cases smaller fans are entirely located inside larger fans). Moreover, some imagery shows darkish features that are clearly not fan shaped, and the model was unable to be trained to agree with human labelers in these cases. We concluded that while humans are readily able to search for particular  CO$_2$ jet seasonal deposits  that conform with a designated shape, this task does not yet align well with the computer vision problems that supervised machine learning excels at. Nevertheless, it is anticipated that new supervised learning methods can be developed that do much better than we were able so far.

For this paper, we focus on predicting the locations or presence of  CO$_2$ jet seasonal deposits in HiRISE images by use of {\em semantic segmentation}~\citep{Long15}. This is a computer vision approach in which different features of an image are delineated, i.e.~segmented, by automatically considering every individual pixel, and classifying it as belonging to exactly one category, out of a set of mutually exclusive categories. In our application in this paper, the semantic segmentation task is binary, since each pixel needs to be categorised as belonging to a seasonal fan or blotch, or belonging to a {\em background} class. In this Section, we introduce the metrics we used for comparing the extent of agreement between  algorithmically generated semantic segmentations of an image with either a ``ground-truth'' or alternative segmentation generated by other methods, in this case the Planet Four catalog. 

\subsection{Metrics for Semantic Segmentation Accuracy}

Since the aim is binary classification of pixels, there are two types of   {\em disagreement}  that can occur when using semantic segmentation. We  emphasise {\em disagreement} here rather than {\em error}, because although we want our methods to predict seasonal fans/blotches at least as well as the Planet Four volunteers, there are known ways in which different human labelers disagree, and some image features in which the presence of a fan or blotch is subjective. We discuss this further later in the paper. Irrespective, because the segmentation task is binary, standard metrics for any binary classification task are relevant. It is worth noting that some of the metrics we use have previously been used for another citizen science project, in which craters were identified on Mars~\citep{Sprinks19}. Here, we use recall (also known as sensitivity), precision, specificity and  balanced accuracy; these are defined in terms of True Positives (TP), False Positives (FP), False Negatives (FN) and True Negatives (TN) as follows.

We have
\begin{equation}\label{recall}
{\rm Recall} = \frac{{\rm TP}}{{\rm TP + FN}},
\end{equation}
\begin{equation}\label{precision}
{\rm Precision} = \frac{{\rm TP}}{{\rm TP + FP}},
\end{equation}
\begin{equation}\label{specificity}
{\rm Specificity} = \frac{{\rm TN}}{{\rm TN + FP}},
\end{equation}
and
\begin{equation}\label{balacc}
{\rm Balanced~Accuracy} = \frac{{\rm Recall}+{\rm Specificity}}{2}.
\end{equation}
Balanced accuracy is preferred to overall accuracy when the number of samples in each class is unequal, as is the case for our data, for which the two classes are ``background'' and ``CO$_2$ jet seasonal deposits.''

\begin{figure}
\centering
\includegraphics[width=\textwidth]{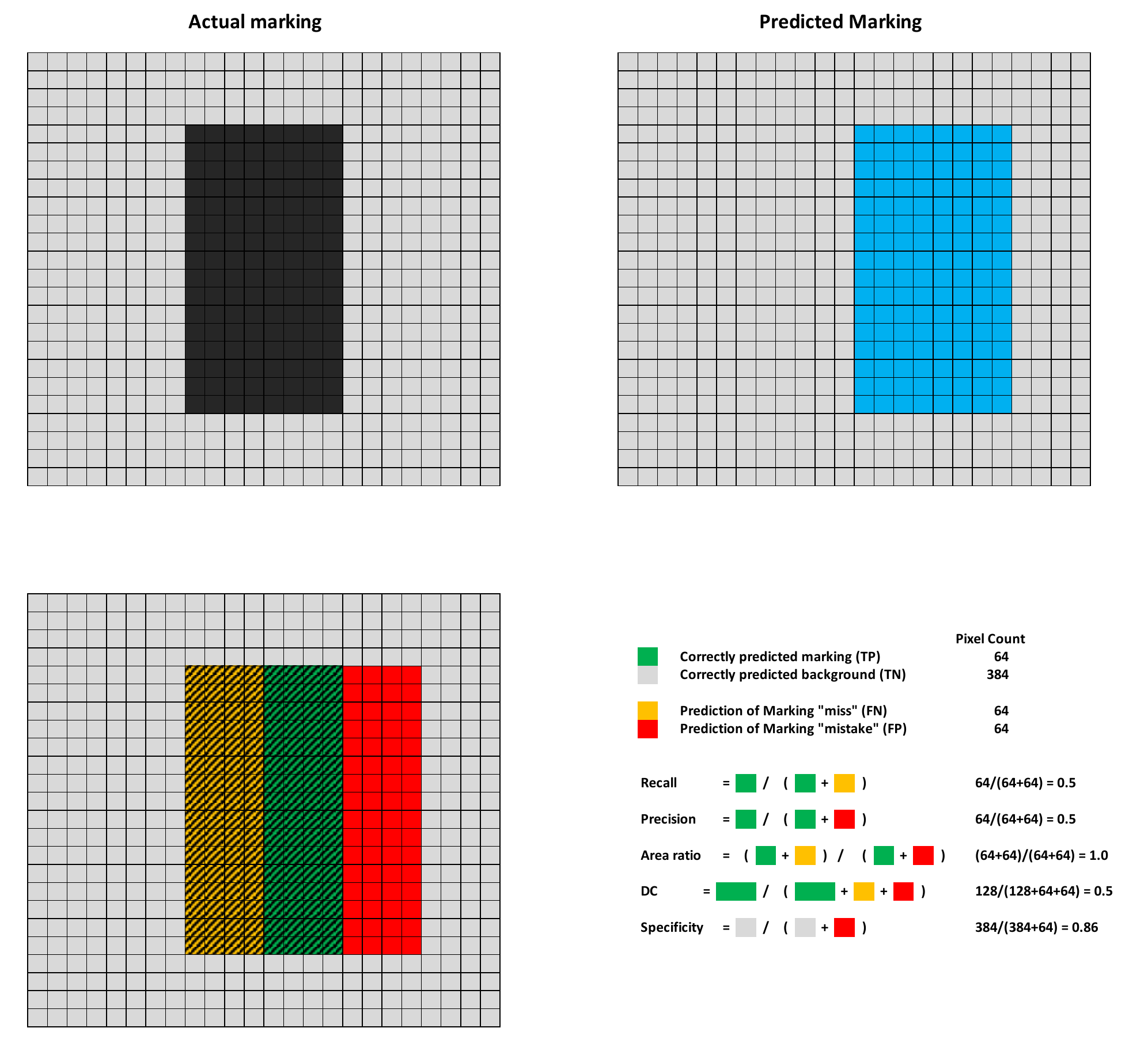}
\caption{Example 1 of Dice Coefficient (DC) with value of 0.5. ``Actual marking'' refers to the Planet Four catalog. ``Predicted marking'' refers to the output of an algorithm and is exactly misaligned by half of the actual marking. Cross hatching in the colour-coded lower figure indicates the position of the ``Actual marking.'' The larger green box in the DC equation illustrates that Dice Coefficient weights True Positives by a factor of 2.}
\label{fig:JI_ex_1}
\end{figure}

\begin{figure}
\centering
\includegraphics[width=\textwidth]{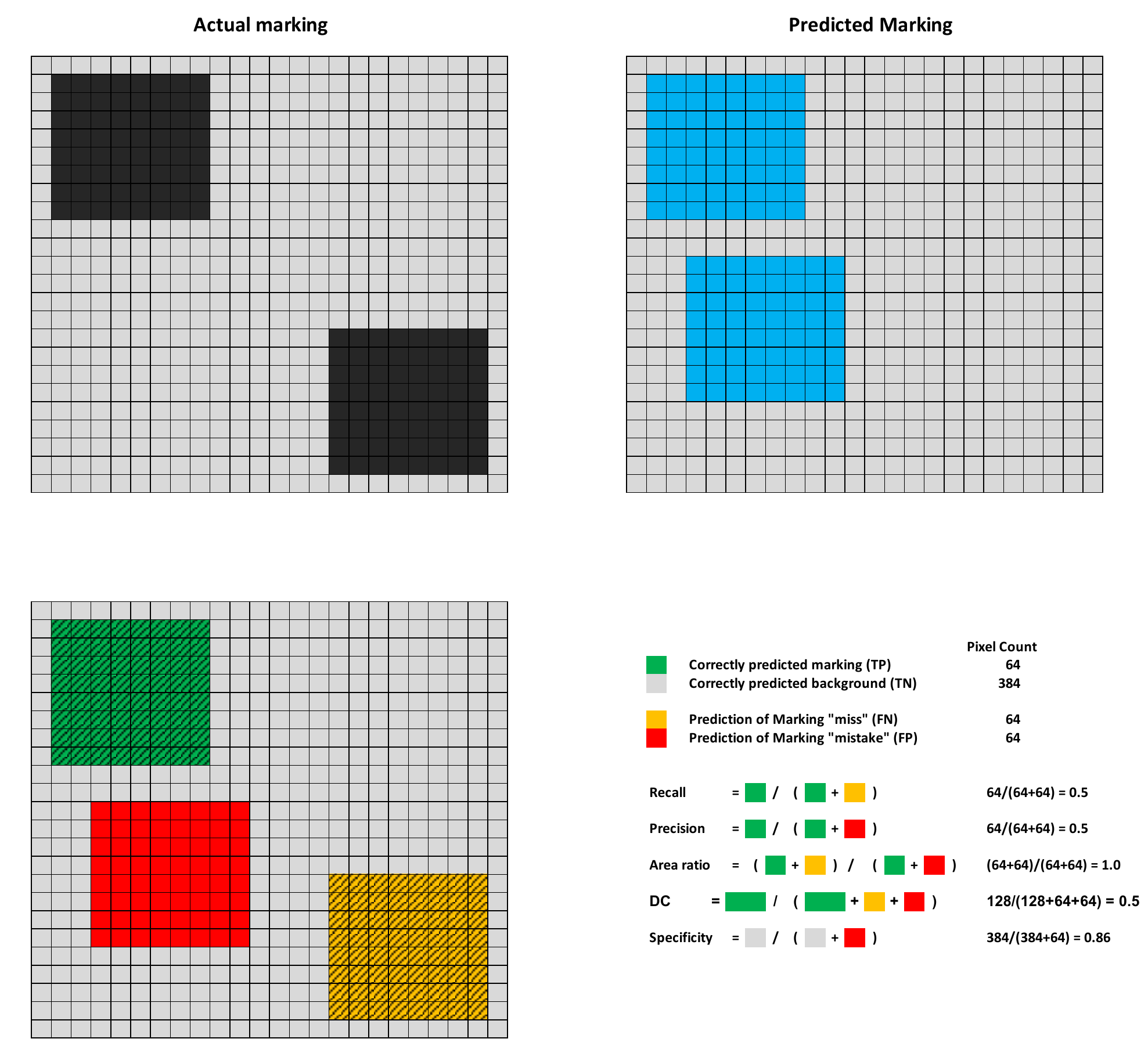}
\caption{Example 2 of Dice Coefficient (DC) with value of 0.5. ``Actual marking'' refers to the Planet Four catalog. ``Predicted marking'' refers to the output of an algorithm. Cross hatching in the colour-coded lower figure indicates the position of the ``Actual marking.'' Unlike Figure~\ref{fig:JI_ex_1} where the prediction is partially aligned with the actual marking, this figure shows the two extreme cases of disagreement, i.e.~the cases where an entire actual marking is missed in the predictions, and where a marking not present at all in the actuals is predicted. The larger green box in the DC equation illustrates that Dice Coefficient weights True Positives by a factor of 2.}
\label{fig:JI_ex_2}
\end{figure}

In our context, recall can be thought of as measuring the fraction of Planet Four catalog fan or blotch pixels that our semantic segmentation models agreed with, whereas precision can be thought of as the fraction of pixels the models predicted to be  CO$_2$ jet seasonal deposits, that were also labeled as fans or blotches in the Planet Four catalog. Specificity measures the fraction of pixels not in the Planet Four catalog that were also not predicted as  CO$_2$ jet seasonal deposits by the models.  Figure~\ref{fig:JI_ex_1} and~\ref{fig:JI_ex_2} illustrate two typical situations that result in precision and recall each having values of 0.5, for different reasons, as explained in the caption of Figure~\ref{fig:JI_ex_2}.

Semantic segmentation methods are also typically assessed using single number metrics that take into account both recall and precision, in the sense that both need to be high for the metric to be high. One commonly used metric is known as Jaccard Index (JI) or IOU (Intersection Over Union)~\citep{Rahman16}; it measures the amount of overlap between two areas, as given by the intersection between the two areas divided by their union. As described later, JI is relevant for how we train our CNN.  However, for measuring semantic segmentation performance, we use another popular metric, the harmonic mean of precision and recall, also known as the Dice Coefficient (DC), or F1-score~\citep{jadon2020survey}.  The reason for preferring DC is that it tends to have a numeric value comparable to both recall and precision when both are similar.

Formally, DC must be defined relative to a designated `positive' class. In this case, the positive class is pixels that are included in fan/blotch markings in the Planet Four catalog. Hence, DC can be expressed as the ratio of twice the number of correctly predicted pixels that are in fan/blotch markings in the Planet Four catalog, TP, to the sum of the pixels in fans/blotches in the Planet Four catalog and those predicted by the algorithm to be in fans/blotches. The latter is equal to  the sum of 2TP, FP and FN. Hence, 
\begin{equation}\label{JI}
{\rm DC} = \frac{{\rm 2 \cdot TP}}{{\rm 2\cdot TP + FP + FN}}.
\end{equation}
To illustrate, Figures~\ref{fig:JI_ex_1} and~\ref{fig:JI_ex_2} show two examples where TP=FP=FN, resulting in DC$=0.5$. In Figure~\ref{fig:JI_ex_1}, the prediction is misaligned with the actual area by half the actual marking area's width. In Figure~\ref{fig:JI_ex_2}, an actual marking is correctly predicted for all pixels, another is entirely missed, and a mistakenly predicted area of the same size produces false positives. We also use 
\begin{equation}\label{area_ratio}
{\rm log~area~ratio} = \log{\left(\frac{{\rm TP+FN}}{{\rm TP + FP}}\right)},
\end{equation}
i.e.~the log ratio of the total number of pixels in the positive class according to the Planet Four catalog to the total number of pixels predicted by a model to be in the positive class. A positive value of this metric means that less total area of fans and blotches was predicted by the model than by the Planet Four catalog; a negative value is the opposite. This metric can, by itself, be misleading, since it can have a perfect value of 0.0 despite no overlap in actuals and predictions; hence log-area-ratio is useful only when considered in combination with a metric that takes into account both recall and precision, such as Dice coefficient.

The final metric we consider is one that measures whether our semantic segmentation methods agree with the Planet Four catalog as to whether the center pixel of Planet Four catalog fans or blotches should be predicted as a  CO$_2$ jet seasonal deposits.  We call this metric {Center-Overlap}; it is defined analogously to {\em recall}, i.e.~it is the fraction of Planet Four catalog fan or blotch shape center pixels (`P4 Centers' -- PC) that our algorithms correctly predicted (`Matching Centers'  -- MC) as belonging to a marking. This can be expressed as
\begin{equation}\label{centre_overlap}
{\rm Center~Overlap} = \frac{{\rm MC}}{{\rm PC}}.
\end{equation}
The value of Center-Overlap is that it provides an indication of whether the machine learning method agrees with the Planet Four catalog on whether a marking should be present where a human says it should be. 

\subsection{Masks for Planet Four Catalog Markings}

For comparison of all model results with the Planet Four catalog, we constructed binary mask images of the same size as the original HiRISE images by constructing filled polygons corresponding to the pixels interior to the parameterized cones and ellipses in the P4 catalog designating each fan and blotch marking. Many fans and blotches overlapped. As our primary machine learning aim was binary image segmentation, we ignored whether polygons were blotches or fans, and simply set a pixel value to one in the mask if it was contained in any fan or blotch. See Figure~\ref{fig:feature_masks_on_HiRISE} (center-left of each example) and Section~\ref{S:3} for example mask images.  Its worth noting that binary statistics have been used before for citizen science projects regarding Mars, namely identification of craters~\citep{Sprinks19}. 

\section{ISODATA Clustering Baseline}~\label{S:ISODATA}

To motivate the need for a CNN and validate the effectiveness of our CNN design we sought baseline results using a  traditional method for semantic segmentation, namely ISODATA~\citep{Ball}. This is an unsupervised clustering algorithm similar to, but more advanced than, k-means clustering. ISODATA is widely used in multi- and hyperspectral satellite remote sensing for the separation of spectral features and is applied to feature detection and mapping tasks across a wide arrray of applications, including detection of ecosystem degradation \citep{Abdollahzadeh2021}, mineral mapping \citep{Mahboob2019}, and weed detection \citep{Stroppiana2019}.  In our current context, ISODATA performs an unsupervised clustering and classification of the 3 spectral band HiRISE images, and then the cluster with the lowest red brightness is used as the potential feature detection and provides a segmentation of the darkest features in the HiRISE images. The aim in ISODATA is to approximate the natural structure of a multidimensional dataset by iteratively passing it over the data and defining classes by minimizing pixel separation values. The detection of features is independent of the spatial relationships between pixels, and does not incorporate any {\em a priori} knowledge of the spectral character of the features (other than that they are darker than the landscape). The ISODATA algorithm makes no assumptions on the underlying probability distribution of the datasets. By selecting the darkest class partitioned by ISODATA, the output is a binary prediction for whether each individual pixel is a part of a CO$_2$ jet seasonal deposit, or part of background.  Therefore ISODATA in this application can be considered to produce semantic segmentation, just like our CNN does.

\subsection{Design}

The automated ISODATA spectral clustering was implemented through an IDL (Interactive Data Language) programmatic workflow~\citep{Didriksen}, utilising ENVI (Environment for Visualizing Images) software custom tasks~\citep{envitask}.  As ISODATA  clustering is based on the underlying statistics of the image, no training data is required.  The following parameters can be tuned to adjust the performance of the ISODATA algorithm: number of iterations (I), number of classes (N) and the convergence threshold (C).  The number of iterations was set high enough that it became redundant and did not determine the exit condition of the algorithm (I = 200).  Instead, the algorithm exit was determined by the convergence threshold, which was set to 99.99\% and was achieved on all images.  A range of values of N were tested to gauge the sensitivity of the clustering to this parameter.  A slight  performance improvement in feature segmentation was observed with increasing N from 2 to 5. % (see Figure~\ref{fig:ISO1}).
In particular, the number of false positives was slightly reduced in images where a substantial fraction of non-feature dark pixels existed, for example, shadows from topographic variation, and so-called ``spider channels''~\citep{piqeux2003}.  When assessed over a large number of images however, values of N$>$5 provided no substantial improvement in accuracy metrics, but substantially increased computational time.  A value of N=5 was selected to balance computational time, with spectral sensitivity.  The class with the lowest brightness values (darkest in the red band) was then selected and the 5 classes collapsed to 2, with the darkest class representing the potential feature detection and pixels belonging to all other classes representing background.  This provided a binary segmentation for direct comparison with the Planet Four masks and the output of the CNN.     Basic post-processing cleaning of the ISODATA clustering output was undertaken to remove small isolated areas of false-positives which were observed particularly in HiRISE images with a poor signal-to-noise ratio. A region size threshold (R) was utilised to remove all isolated areas below a certain size (number of pixels) through application of the 'label region' IDL code \citep{envitasklabel}.  This procedure consecutively labels all of the regions, or distinct groups, in a binary image with a unique region index, allowing the number of features and their size (number of pixels) to be determined.  A range of R values were tested, % (Figure~\ref{fig:ISO2}),
with a final selection of R=150 chosen to optimally balance (i) minimization of false positives, which were higher in images of poor signal-to-noise, small scale elevation change (resulting in shadow pixels), and featureless images; and (ii) minimization of misses (real features that aren't detected).  Thus all feature areas smaller than 150 contiguous pixels were removed from the feature detection class and assigned to the background class.  All images shown have had the small feature filtering applied.  Unlike supervised learning, it is not necessary to use a cross-validation method for ISODATA. Instead, the same algorithm is applied to all images, and the results reported.

%In general the number of false positives is minimized by increasing N (number of classes) and increasing R (cleaning threshold determining region size), while these same actions can increases the number of misses on some images. 

%\begin{figure}[h]
%\centering
%\includegraphics[width=5cm]{mars-viking-zoom.jpg}
%\caption{Will show examples of feature labeling before and after cleaning, and with different values of the change threshold and number of classes.}
%\label{fig:example_feature_labeling}
%\end{figure}  

\subsection{Results}

We report results from ISODATA in Figures~\ref{fig:Res0}--\ref{fig:Res4} and Figure~\ref{fig:Res5}, alongside results from our trained CNNs (described in the following Section).  As can be expected from the known effectiveness of deep CNNs for semantic segmentation~\citep{Minaee21,Yuan21}, we generally achieve better results using CNNs. Regardless of the performance of ISODATA clustering when compared with the CNN, the method has clear limitations.  ISODATA relies on manual parameter selection to identify which spectral class relates to the feature of interest.  Thus the model developed is specific to the problem of identifying spectral features darker than image background, and would not be generalizable to a different image segmentation context.  Generalization capability is, however, a strength of CNN architectures.

\section{Supervised Learning for Semantic Segmentation}\label{method-ml}

We now present our primary approach to using semantic segmentation to  identify  CO$_2$ jet seasonal deposits on HiRISE images, i.e.~deep CNNs trained by supervised learning. Such use of CNNs for semantic segmentation has proliferated for both remote sensing~\citep{Yuan21} and more generally~\citep{Minaee21}, following the introduction of the U-net CNN form in 2015~\citep{Ronneberger15}.  Regardless of exact architecture, CNNs are ideally suited to taking into account the context of other pixels in an image when classifying each individual pixel.  The specific form of CNN we use is a very recently developed CNN called a HRNet~\citep{Wang.20}.  

\subsection{Design}

\subsubsection{Training and Validation Design}

In order to ensure that we trained a model with maximum generalizability, and that did not overfit to the physical appearance of a specific location on Mars, we used a version of Leave-One-Out (LOO) cross-validation~\citep{LOO}. The idea in LOO is to train a model on all $N$ available samples apart from one (which is `left out') and then test the model's performance on that remaining sample. The process is then repeated such that every sample is left out once, resulting in $N$ models being trained. 

We chose to treat each polar Mars region (see Section~\ref{S:HiRISE}) as a sample, i.e. $N=28$. The reason for this was twofold. First, this choice was expected to produce a more robust machine learning model than, for example, treating each individual image as a sample, whereby because spatially nearby images are expected to be more similar than those further apart, a left-out image for model testing would often be similar to images used during training. Hence test performance would not be as indicative of actual model generalizability.  Second, an even more important situation to avoid was splitting of images that showed the same location into both the training set and a validation set. This situation is relevant  because the HiRISE images we used were acquired in two consecutive Martian years, and there were many cases of image pairs amongst the 221 images that were of the same location from two consecutive years. Given that fans and blotches and their directions may repeat each year, treating each region as a sample in this way ensured that any images of the same location from both years were either both in the training set, or both in the left-out validation region.

\subsubsection{Deep Convolutional Neural Network Architecture and Training}

Our HRNet design closely followed that of~\cite{Wang.20}.
We trained our HRNets (implemented in TensorFlow 2.1) from scratch (i.e.~random initial values for all parameters) using stochastic gradient descent (SGD), (with momentum parameter equal to $0.9$ and weight decay parameter equal to $10^{-4}$ on all weights) and the soft-Jaccard Index loss function (also known as IOU loss)~\citep{Rahman16}. The input data for SGD were patches extracted from the large HiRISE images of size 512$\times$512 pixels, with a total batch size of 20, using 4 parallel GPUs, each with a sub-batch of size 5. This relatively small batch size is required to ensure our GPUs did not run out of RAM when using this large input patch size. Training followed a stepped learning rate schedule for 80 epochs, where one epoch is equal to the number of batches required before all training images are sampled from once without repetition; hence the total number of patches used during training was 80 times the number of training images. The total number of patches learned from during training differs depending on which region is left out, but is in the order of 220 HiRISE images times 80 epochs, i.e.~$\sim17000$ patches for each model. Given the semantic segmentation task we use, this equates to supervised learning from over 4.4 billion pixels. The learning rate was 0.03 for 40 epochs, 0.003 for another 30, and 0.0003 for the final 10 epochs.

The order in which images were used in each epoch was randomly shuffled for each epoch. To ensure robustness of the model to the three map scales, the 20 selected HiRISE images for each batch were independently randomly chosen to be scaled at 25cm, 50cm or 100cm map scale, and if the scale did not match its original scale, resized accordingly using bicubic interpolation. Next, a random location within each image was chosen for cropping of the 512$\times$512 color patch. We used spatial data augmentation such that each tile chosen in a batch was flipped vertically with probability 0.5, horizontally with probability 0.5, and rotated 90 degrees with probability 0.5. To mitigate image-edge effects due to the implicit use of black pixels on borders of images by ``same'' mode convolution operations, before cropping a patch each image was padded with grey pixels. For map scale 100cm, the padding was 16 pixels on all sides; for map scale 50cm, with 32 pixels; and for map scale 25cm with 64 pixels. 

\subsubsection{Inference}

In machine learning, `inference' refers to the application of a trained model to data that was not used during training; learning is disabled at this point.  For evaluation of our trained HRNets, each model was run in inference mode on each image in the corresponding left out region. Our HRNets were specifically designed to enable input patches for inference (which we call ``chunks'') to be larger than patches used for training, as larger patches minimise artifacts due to image-edge effects. We used the maximum chunk size for inference supported by our GPU: 4096$\times$4096. As most images were far less wide than this, we padded with grey pixels on the left and right of the image when necessary. We additionally padded the top and bottom of the images by 128 pixels. This padding matched padding used on image boundaries during training; the intent was that the model would learn that a contiguous gray region is ``image border'', rather than a marking. This was also why our padding was grey pixel values rather than black, since markings tend to be close to black (and much darker than grey pixels) in the HiRISE images. 

Following tiling of each validation image into 4096$\times$4096 chunks, each chunk was passed to the HRNet to predict a binary segmentation mask output. The resulting chunks were concatenated together, and padding removed, to construct a mask of the same size as the original image. Images with tiling that left an unused strip of image down the right hand side then had that strip removed for calculation of metrics. All resulting masks were saved and used to generate the results reported in this paper.

%%%%%%%%%%%%%%%%%%%%%%%%%%%%%%%%%%%%%%%%%%
\subsection{Results}\label{S:3}

\begin{figure}[p]
\centering
%{\includegraphics[width=0.85\linewidth]{ESP_022699_0985_4panel_composite.png}   \label{fig:App1} }
%{\includegraphics[width=0.85\linewidth]{ESP_022379_0930_4panel_composite.png}   \label{fig:App2} }
%{\includegraphics[width=0.85\linewidth]{ESP_020146_0950_4panel_composite.png}   \label{fig:App3} }
%{\includegraphics[width=0.85\linewidth]{ESP_011370_0980_4panel_composite.png}   \label{fig:App4} }
%{\includegraphics[width=0.85\linewidth]{ESP_011351_0945_4panel_composite.png}   \label{fig:App5} }
%{\includegraphics[width=0.85\linewidth]{ESP_011350_0945_4panel_composite.png}   \label{fig:App6} }
%{\includegraphics[width=0.85\linewidth]{ESP_011348_0950_4panel_composite.png}   \label{fig:App7} }
%{\includegraphics[width=0.85\linewidth]{ESP_011341_0980_4panel_composite.png}   \label{fig:App8} }

{\includegraphics[width=0.74\linewidth]{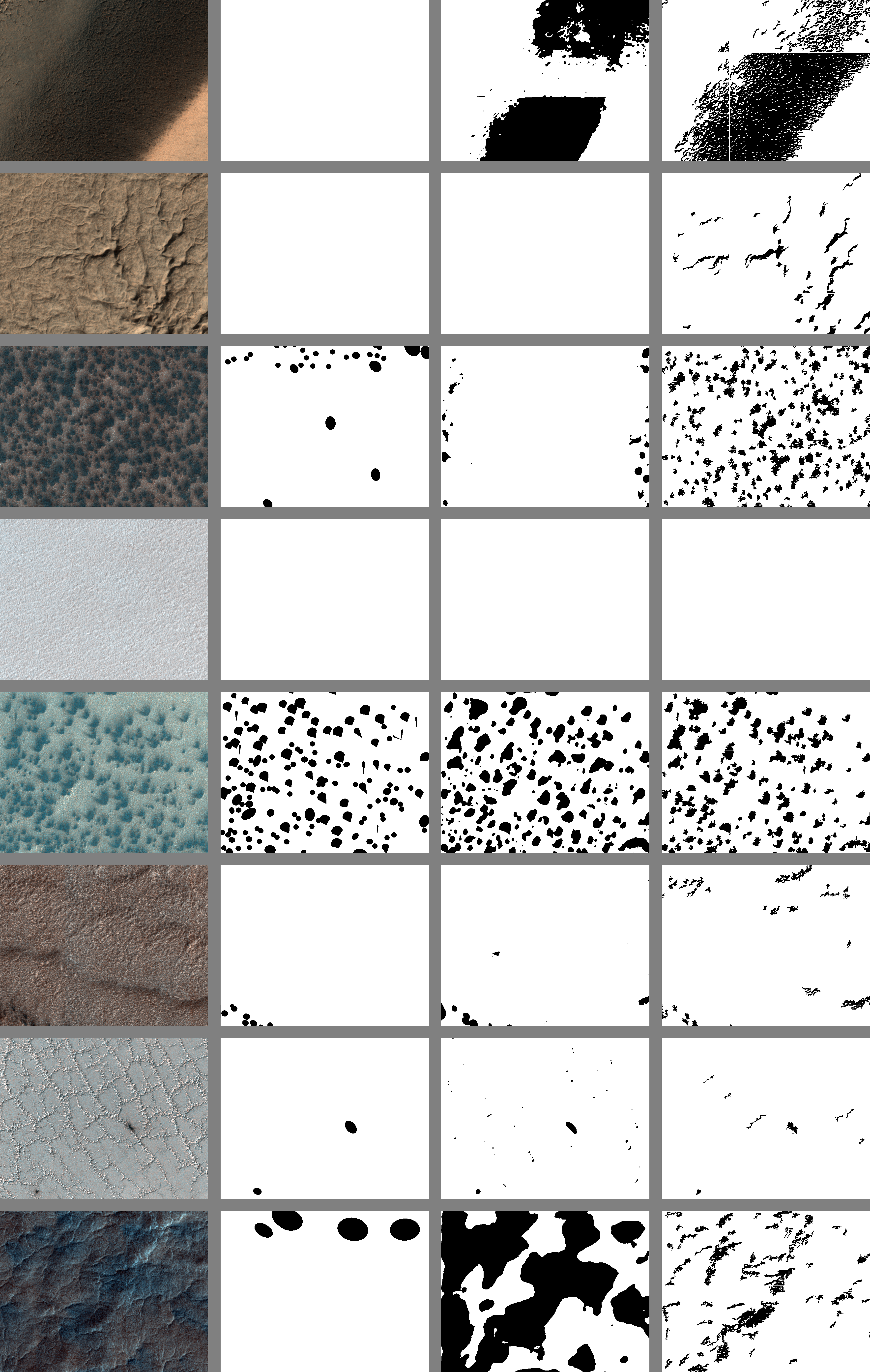}    }\\
{\scriptsize \quad~ A. Planet Four tiles \quad\quad B. Planet Four label masks \quad C. CNN predictions \quad\quad D. ISODATA predictions}
\caption{Each row shows an example of a Planet Four tile cropped from a HiRiSE image (column A), alongside the label masks (column B), created from the Planet Four catalog, the predictions of the trained CNNs from cross-validation  (column C), and the predictions of ISODATA clustering, (column D). From the top, the HiRISE images are ESP\_022699\_0985, ESP\_022379\_0930, ESP\_020146\_0950, ESP\_011370\_0980, ESP\_011351\_0945, ESP\_011350\_0945, ESP\_011348\_0950, ESP\_011341\_0980. The grey bands are not part of the images.}
\label{fig:hirise_vs_p4_vs_ml}
\end{figure}

Figure~\ref{fig:feature_masks_on_HiRISE} shows two example full size  HiRISE images (left quarter of each example) alongside masks created from the Planet Four catalog (center left), cross-validation predictions from the CNN (center right) and predictions from ISODATA clustering (right).  The first example illustrates a case where the CNN matched the Planet Four catalog well, and the second a case where the CNN's weakness is evident: it tends to predict dark pixels when there is no clear ellipses or fan shape. The model's task is harder in cases like this where the overall image is low in brightness and contrast, and/or showing a lot of shadows. Figure~\ref{fig:hirise_vs_p4_vs_ml} (left column) shows example tiles  cropped out of  full-size images, as seen by Planet Four human labelers (680 pixels high $\times$ 840 wide).

\begin{figure}
\centering
{\includegraphics[width=1\linewidth]{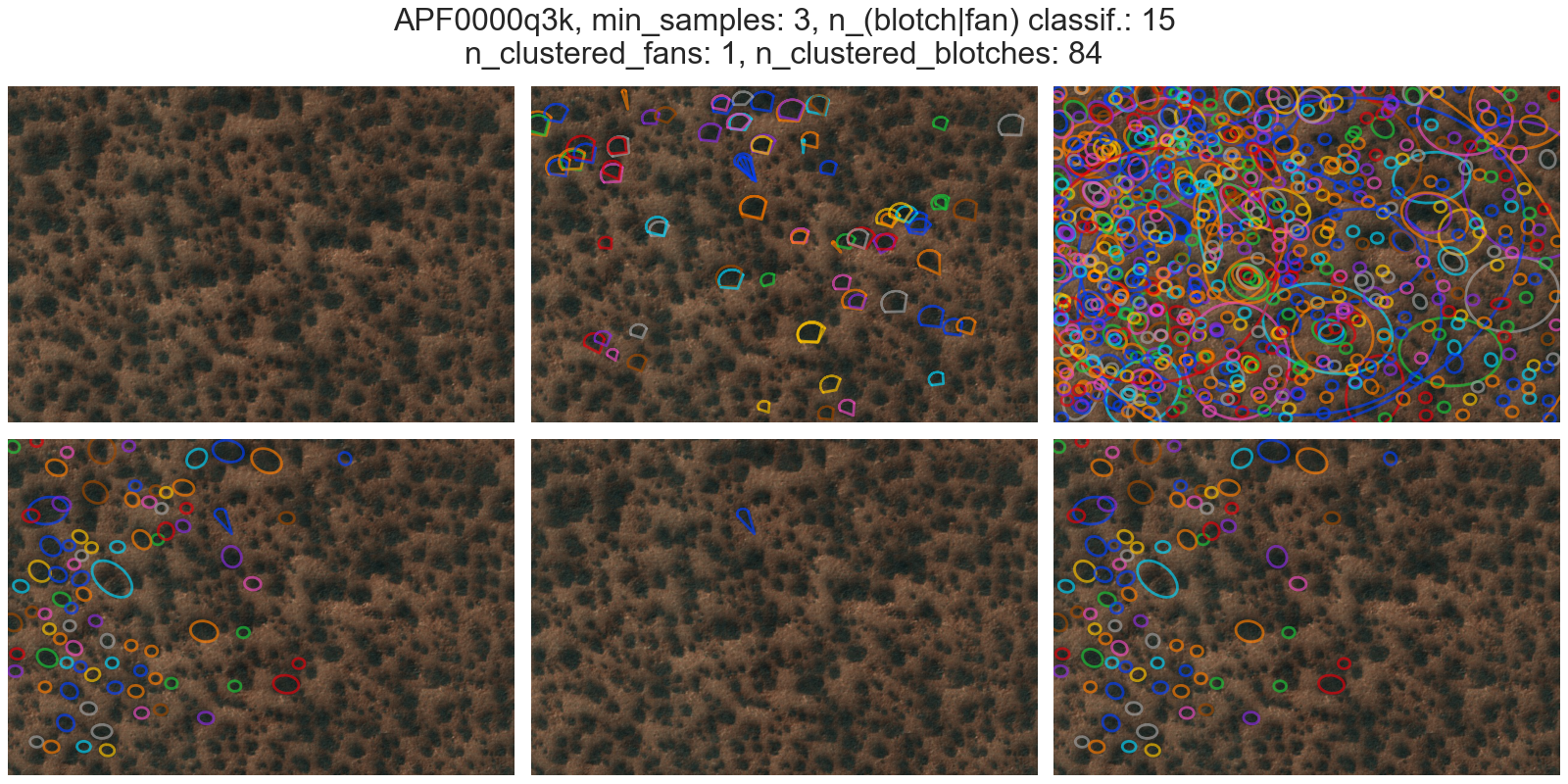}}\\
\caption{Planet Four pipeline for image tile APF0000q3k, using a density-based clustering pipeline as described in \citet{Aye19}.
The \emph{min\_samples} parameter indicates how many markings need to fall within a set of given pixel distances, and it was found that a value of 3 suppresses false positives efficiently, which was the goal of that pipeline.
The parameter \emph{n\_(blotch|fan) classif} indicates how many review submissions actually contained actual markings.
In this case, 15 submissions out of the standard 30 required reviews contained markings, indicating that the complexity of the image tile has made 50\% of the volunteers skip trying 
Upper left: The HiRISE input tile to be marked by volunteers; upper middle: the \emph{fan} markings of 30 volunteers; upper right: the \emph{blotch} markings of 30 volunteers; lower right: resulting blotches after applying density-based clustering and averaging; lower middle: resulting fans after applying density-based clustering and averaging; lower left: the markings entering the catalog after a location-base >50\%-voting between coinciding fan and blotch markings.}
\label{fig:p4_pipeline_fig1}
\end{figure}

One thing to note is the sometimes apparent lack of Planet Four labels, e.g.\ in the third row of Figure~\ref{fig:hirise_vs_p4_vs_ml} or as evident by the two almost empty tiles in the top of Figure~\ref{fig:P4Miss}.
There are two rare instances where the pipeline developed in \citet{Aye19} is under-performing by creating false negatives because it was designed to be trustworthy in terms of preventing false positives.
The first case is when an image tile would contain a large amount of objects.
What happens then is two-fold: a) volunteers either would simply not even start the daunting task of marking so many objects, and b), even the ones that go through with it will have trouble in aligning the markings well with surface features, because they start to overlap a lot, as shown in Figure~\ref{fig:p4_pipeline_fig1}.
In that figure, the parameter \emph{n\_(blotch|fan) classif} being 15 indicates that half of volunteers have not submitted any markings, as the usual retirement requirement was a count of 30 classifications per image tile.
The second scenario where the pipeline seemingly would underperform is when the surface features have not the shape that the volunteers were asked to mark, as it is happening in Figure~\ref{fig:P4Miss}.
There the fans erupt from linear cracks in the seasonal ice and can be 1) very small, making it difficult to be marked by the offered tools, and 2) show rather a ``curtain-like" shape than the ``ice-cone" shape the volunteers were asked to mark.
It was, however, shown in \citet{Aye19} that these incidences are rare, by comparing the volunteers results with a large randomly selected set of example tiles (1\% of all data) that were reviewed by the science team, the so called ``gold data set".

It was always the plan to revisit the efficiency of the Planet Four catalog pipeline \citep{Aye19} by combining image-focused ML techniques with the citizen science based labeling efforts, and we think that this paper is a first step towards that goal. We emphasize that the semantic segmentation method we used is not able to determine fan and blotch shapes; the models are trained to classify individual pixels. It is an open problem in machine learning to address the much harder problem to force a deep CNN to fit image regions to constrained shapes such as ellipses or fans. The remainder of this section quantifies our results overall, by region, by image and by tile.

\subsubsection{Semantic Segmentation Aggregated Over all Images}

Figure~\ref{fig:Res0} provides an overall summary of the performance of the two algorithms we used for semantic segmentation, in comparison with the crowdsourced Planet Four catalog. The comparison is also summarised in Table~\ref{tab:OS}. Data in Figure~\ref{fig:Res0} was produced by summing all pixel-wise TPs, FPs, FNs, FPs, and Center-Overlap over all 221 images, and then calculating aggregate metrics. First, we see from the blue bars that both the CNN and ISODATA clustering methods predict more total area than the Planet Four catalog did (5.7\% coverage), but the ISODATA clustering (11.2\% coverage) overestimates in comparison with Planet Four by a substantially greater amount than the CNN (6.8\% coverage). These findings alone do not enable a conclusion to be reached about which method has better agreement with the Planet Four catalog. However, Figure~\ref{fig:Res0} shows that the CNN had overall higher Dice Coefficient, Recall, Precision, Center-Overlap and specificity than ISODATA clustering, from which we conclude that the CNN is substantially more accurate  than ISODATA clustering. Both methods had a higher recall than precision, consistent with predicting more area than in the Planet Four catalog.

\subsubsection{Semantic Segmentation by Region, Image and Tile}

\begin{figure}
\centering
\includegraphics[width=1\textwidth]{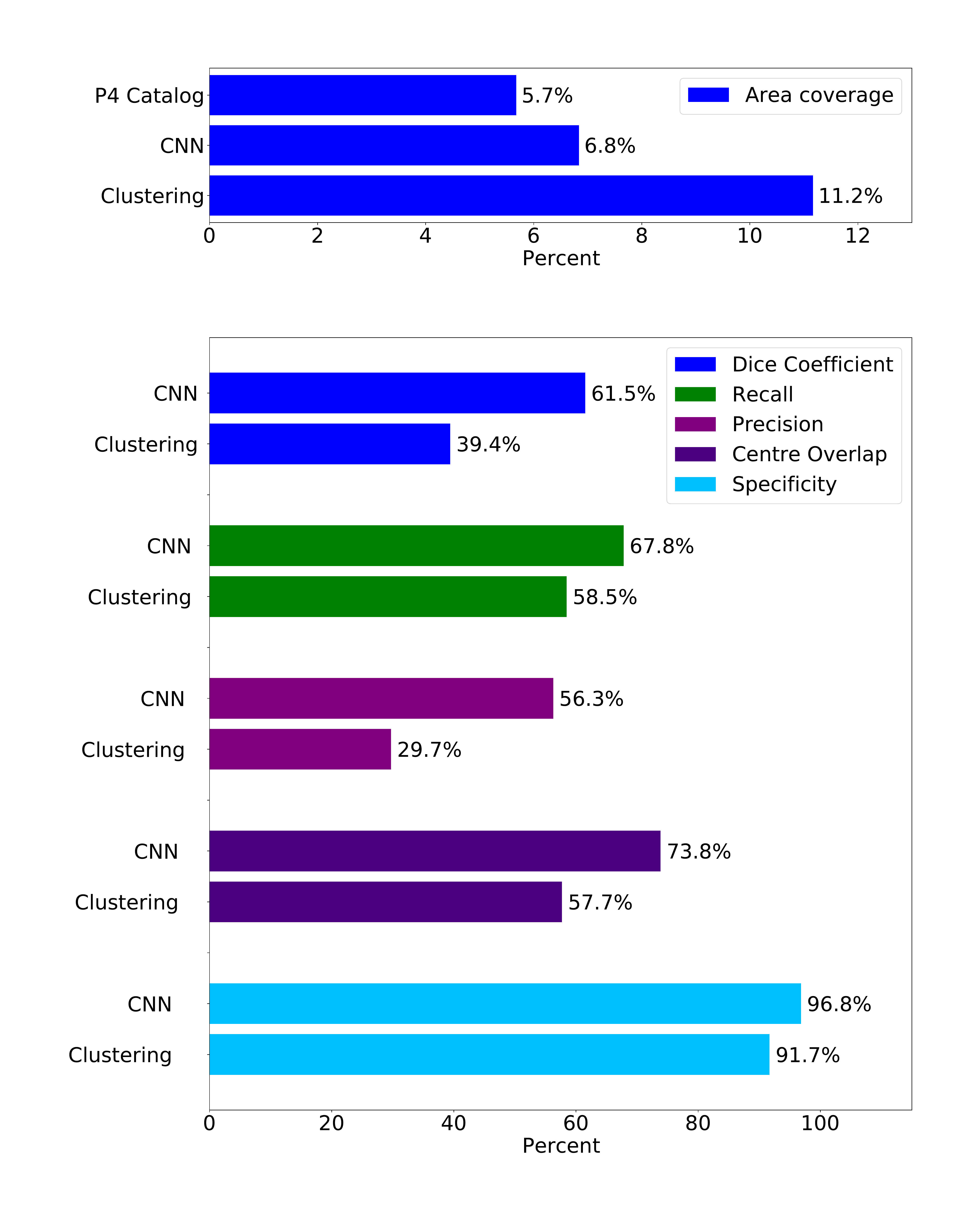}
\caption{{\bf Overall summary.} Summary metrics for area coverage of the Planet Four catalog  compared with the CNN and ISODATA clustering (upper) and summary performance metrics (each metric treats the Planet Four catalog as ``truth'' data) for the two algorithms (lower). The statistics in this figure are aggregated over the entire dataset, i.e.~not on a per-region or per-image basis.}
\label{fig:Res0}
\end{figure}

\begin{figure}
\centering
\includegraphics[width=1\textwidth]{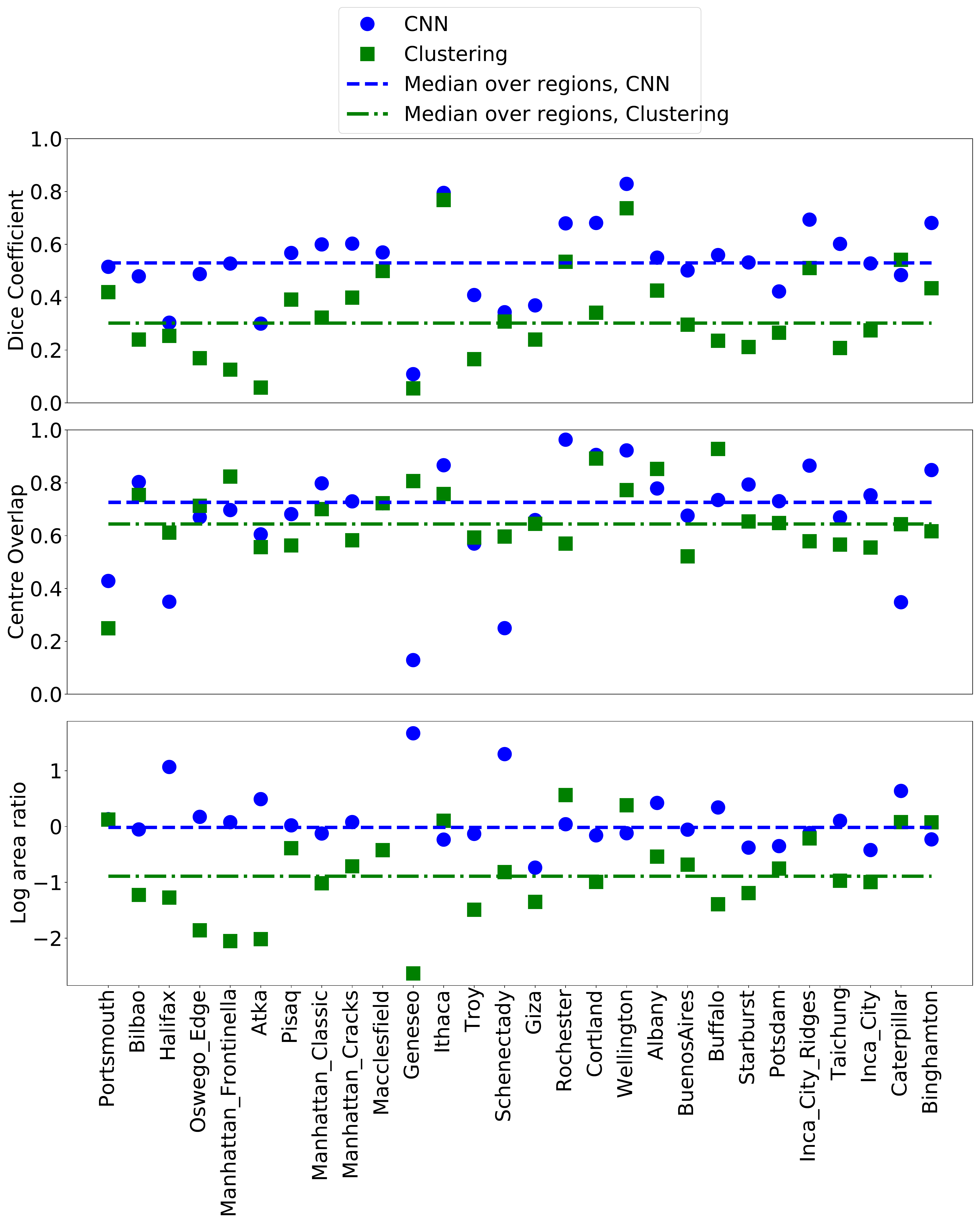}
\caption{{\bf Per-region: Performance of CNN and ISODATA Clustering methods.}  All HiRISE images for a region were aggregated and metrics calculated from the aggregated pixels. The ordering of regions on the x-axis is by increasing latitude for each region.}
\label{fig:Res1}
\end{figure}

\begin{figure}
\centering
\includegraphics[width=0.7\textwidth]{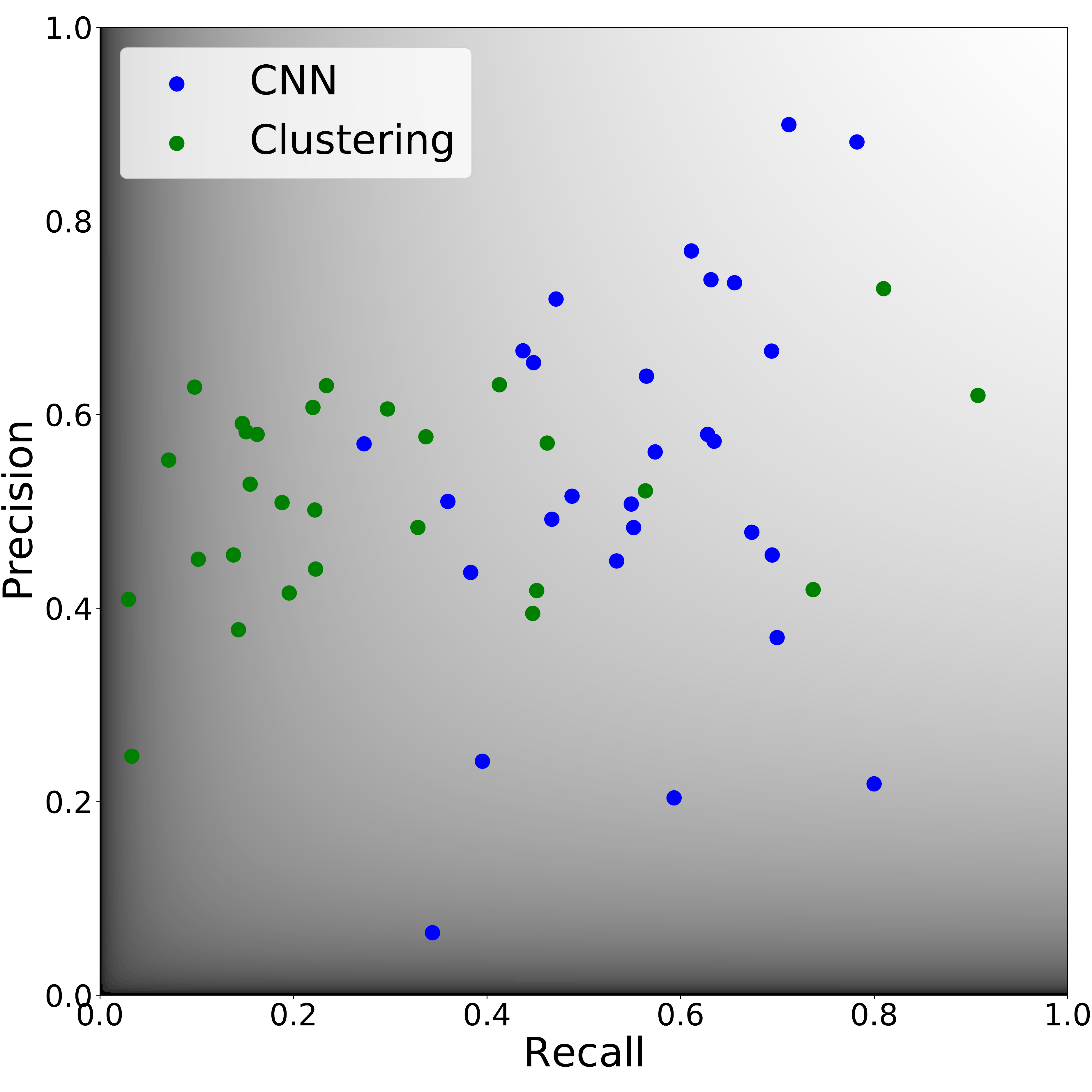}
\caption{{\bf Per-region: tradeoff between recall and precision.} Each marker is a result from one region (all HiRISE images for a region were aggregated and metrics calculate from the aggregated pixels). The background shading lightness represents the value of the Dice Coefficient corresponding to each Recall-Precision pair defined.}
\label{fig:Res2}
\end{figure}

\begin{figure}
\centering
\includegraphics[width=0.99\textwidth]{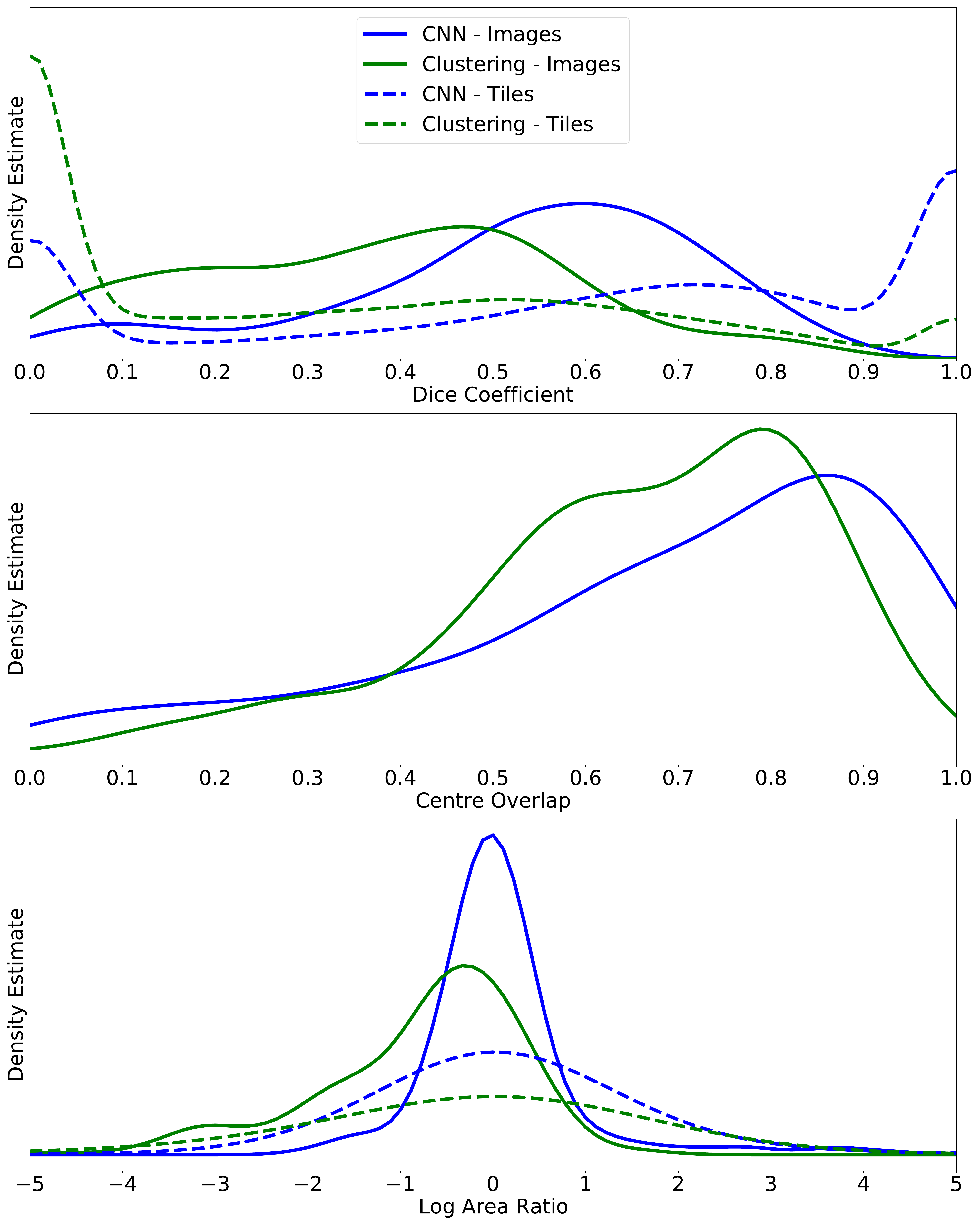}
\caption{{\bf Per-HiRISE-image and Per-tile: Performance of CNN and ISODATA Clustering.} 
The upper subplot shows density estimates of the per-image and per-tile Dice coefficient. The middle subplot shows density estimates for the Center-Overlap for per-image only (due to the relatively large number of tiles with no markings) and the lower subplot shows density estimates for log area ratio.  }
\label{fig:Res3}
\end{figure}

\begin{figure}
\centering
\includegraphics[width=0.7\textwidth]{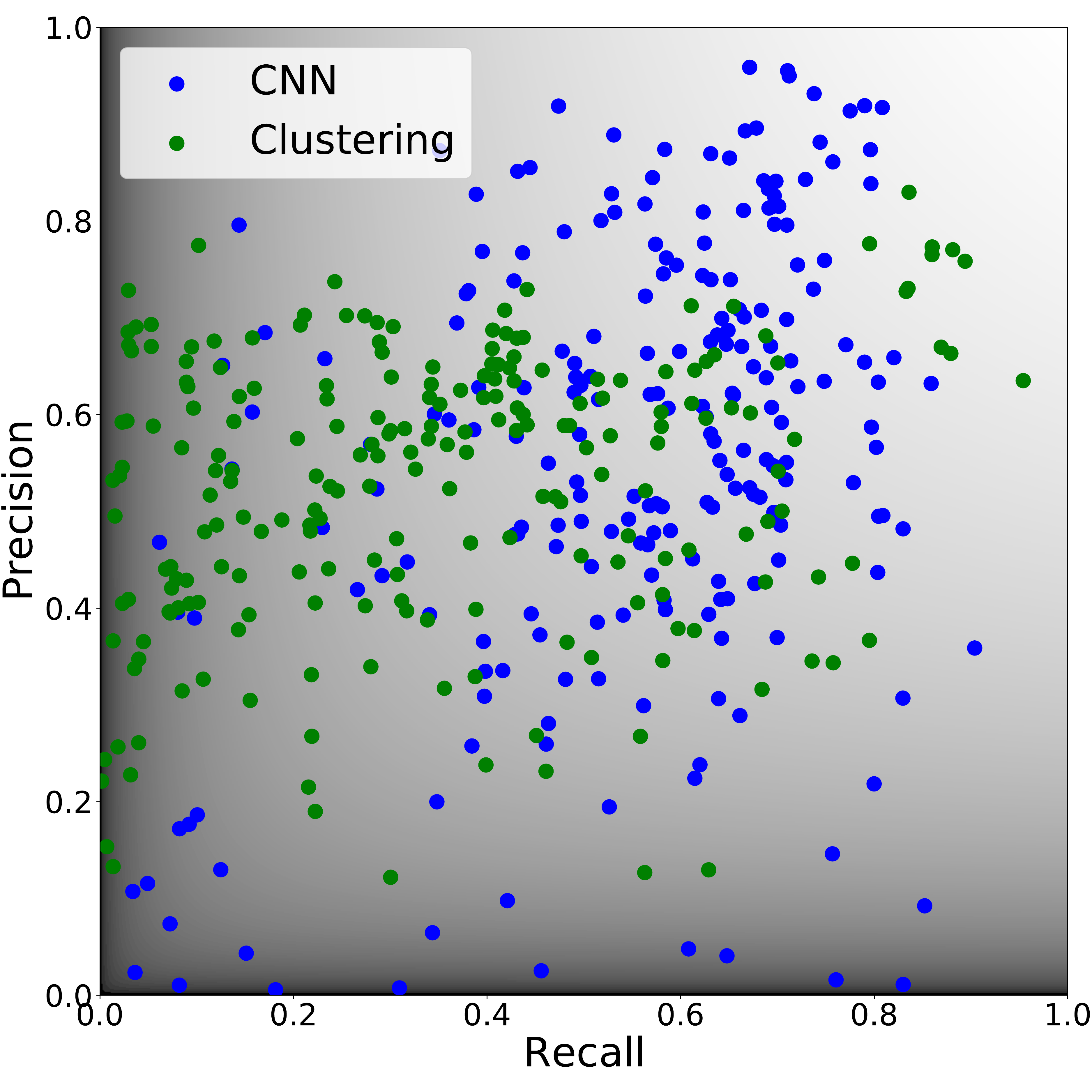}
\caption{{\bf Per-HiRISE-image: tradeoff between recall and precision.} Each marker is a result from one image. The background shading lightness represents the value of the Dice Coefficient corresponding to each Recall-Precision pair defined.}
\label{fig:Res4}
\end{figure}

These overall trends do not hold for all individual regions or images, due to substantial heterogeneity in topography, both between regions, and locally within them. For the machine learning approach, such diversity makes it likely that certain patterns of image features represent outliers that are very difficult to model using supervised learning. For example, as now illustrated, in some cases precision outperforms recall, and in some cases ISODATA clustering outperforms the CNN. Figure~\ref{fig:Res1} shows metrics broken down to aggregates over regions. The upper subplot shows that the CNN method outperforms ISODATA clustering in all but one region, with a median value of 0.55 compared with 0.37. The middle subplot shows that the CNN method also outperforms ISODATA clustering in Center-Overlap overall, but only in 17 of 28 regions. However, this is in part likely to be because ISODATA clustering is much lower in its precision than the CNN, as shown in Figure~\ref{fig:Res0}, meaning that it tends to falsely predict pixels in the  CO$_2$ jet seasonal deposits  class more than the CNN. The lower subplot of Figure~\ref{fig:Res1} shows that the predicted area of the CNN method typically is very close to the Planet Four catalog predictions, whereas ISODATA clustering  frequently predicts significantly greater area, consistent with the trend shown by blue bars in Figure~\ref{fig:Res0}. Figure~\ref{fig:Res2} indicates that for ISODATA clustering, recall varies much more than precision. For many regions, recall is less than 0.3, suggesting that ISODATA clustering has a greater tendency to omit predictions of pixels in the Planet Four catalog than did supervised machine learning. In order to highlight the impact of more localized topography differences, Figure~\ref{fig:Res3} illustrates our results broken down to per-image and per-tile metrics, and Figure~\ref{fig:Res4} on a per-image basis. Figure~\ref{fig:Res3} illustrates again that ISODATA clustering  has greater variability in area ratio, and with many more images and tiles with negative values, meaning ISODATA  produces many more tiles with high numbers of FPs than the CNN.   This is consistent with the coverage shown for ISODATA  in comparison with the Planet Four catalog and the CNN in Figure~\ref{fig:Res0} (top). Figure~\ref{fig:Res3} also shows that variation in Dice Coefficient is more pronounced on a tile basis than an image basis. 

\begin{figure}
\centering
\includegraphics[width=1\textwidth]{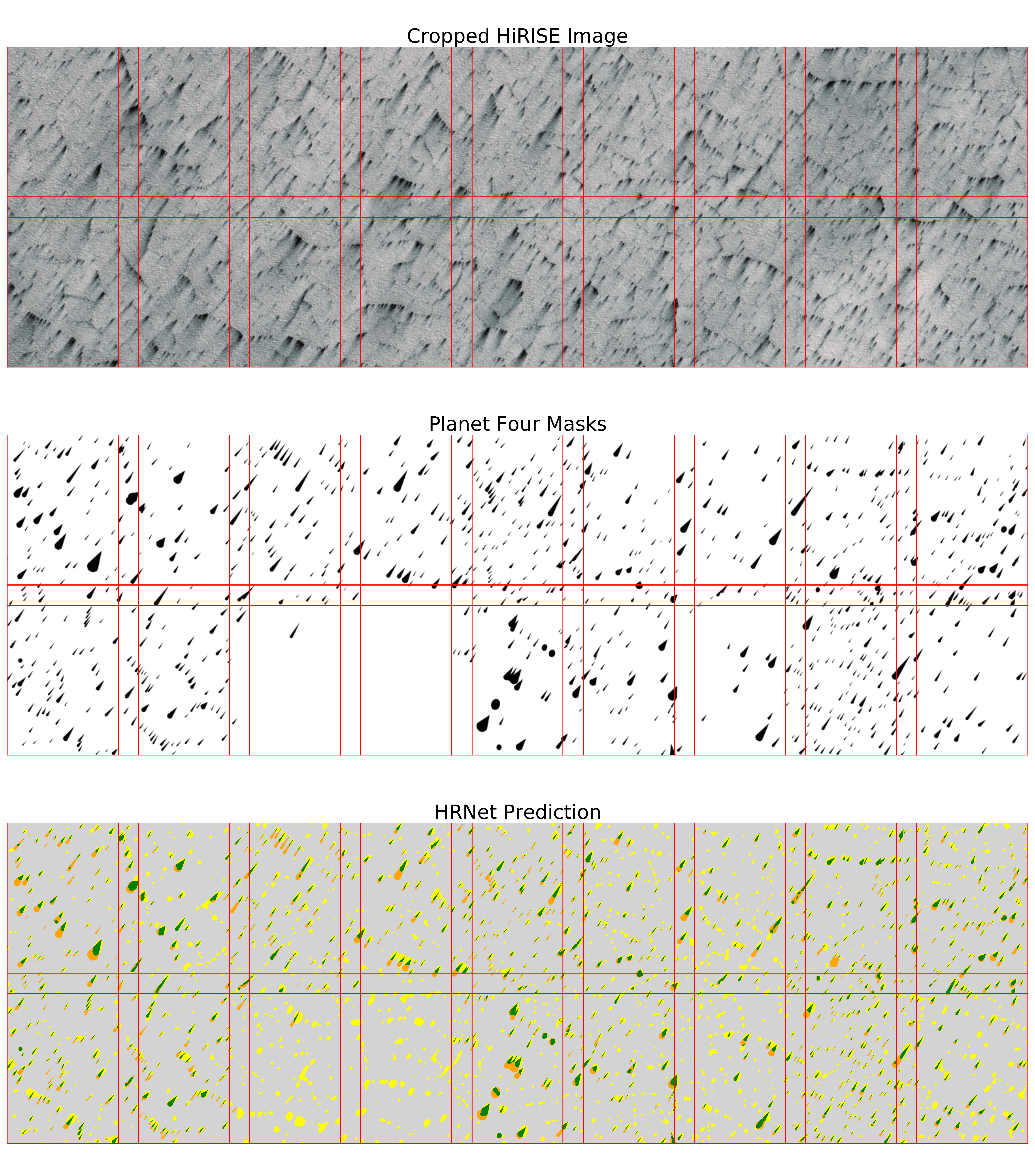}
\caption{{\bf One source of error.}  This figure shows 18 of the overlapping Planet Four tiles cropped from HiRISE Image ESP\_012461\_0925 (top), the corresponding masks we created from the Planet Four catalog (center) and the cross-validation results from our HRNet applied to the HiRISE image (bottom). In the HRNet image, green indicates true positives, yellow false positives and orange false negatives. Red lines indicate the tile boundaries---note the 100 pixel overlap. This data suggests that two of the tiles shown to HiRISE volunteers did not receive enough, if any, markings. Consequently, this will have reduced the Dice Coefficient and other metrics of HRNet success on this image, due to the false positives caused by missed labels.}
\label{fig:P4Miss}
\end{figure}

Finally, Figure~\ref{fig:P4Miss} shows an example of how our metrics of performance are affected by inconsistencies in the Planet Four catalog. The figure shows a case where two tiles within a HiRISE image appear to have not been reviewed by sufficiently many human volunteers. As discussed in Section~\ref{S:4}, one application of our machine learning model could be to help identify such occurrences in human image labeling.

\clearpage

\begin{table}[]
    \centering
    \begin{tabular}{|c|c|c|}
    \hline
         & CNN   & Clustering  \\
         \hline
   Area coverage      &6.8\% &11.2\%\\
         \hline
   Dice Coefficient      &61.5\% &39.4\%\\
   \hline
   Recall      &67.8\% &58.5\%\\
   \hline
   Precision      &56.3\% &29.7\%\\
   \hline
   Centre Overlap      &73.8\% &57.7\%\\
   \hline
   Specificity      &96.8\% &91.7\%\\
   \hline
    \end{tabular}
    \caption{{\bf Overall Summary in Table Form.} The area coverage for the Planet Four Catalog is omitted; see Figure~\ref{fig:Res0} (top). }
    \label{tab:OS}
\end{table}

\clearpage

\subsubsection{Impact of solar longitude on Semantic Segmentation Results}

\begin{figure}
\centering
\includegraphics[width=0.45\textwidth]{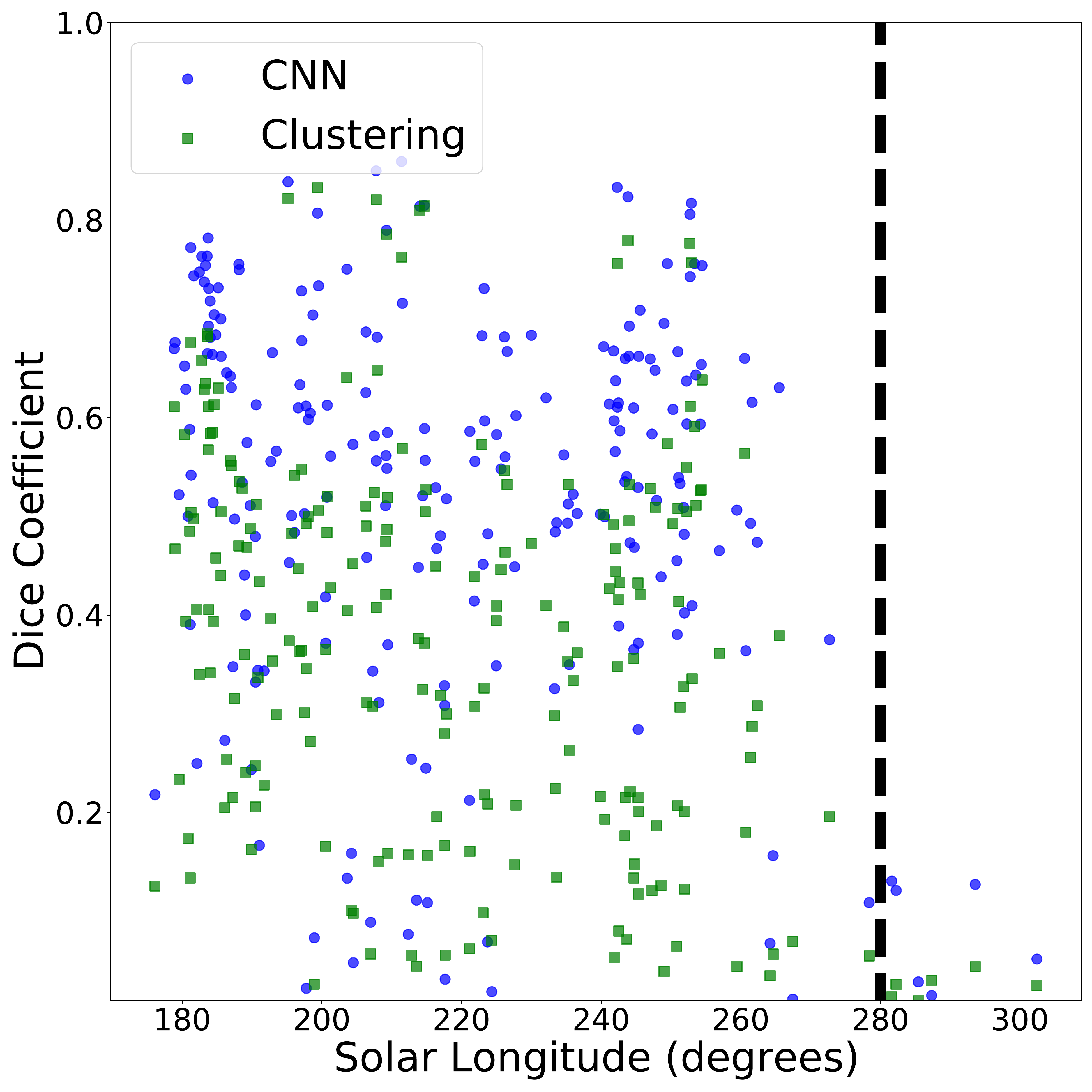}\quad\quad
\includegraphics[width=0.45\textwidth]{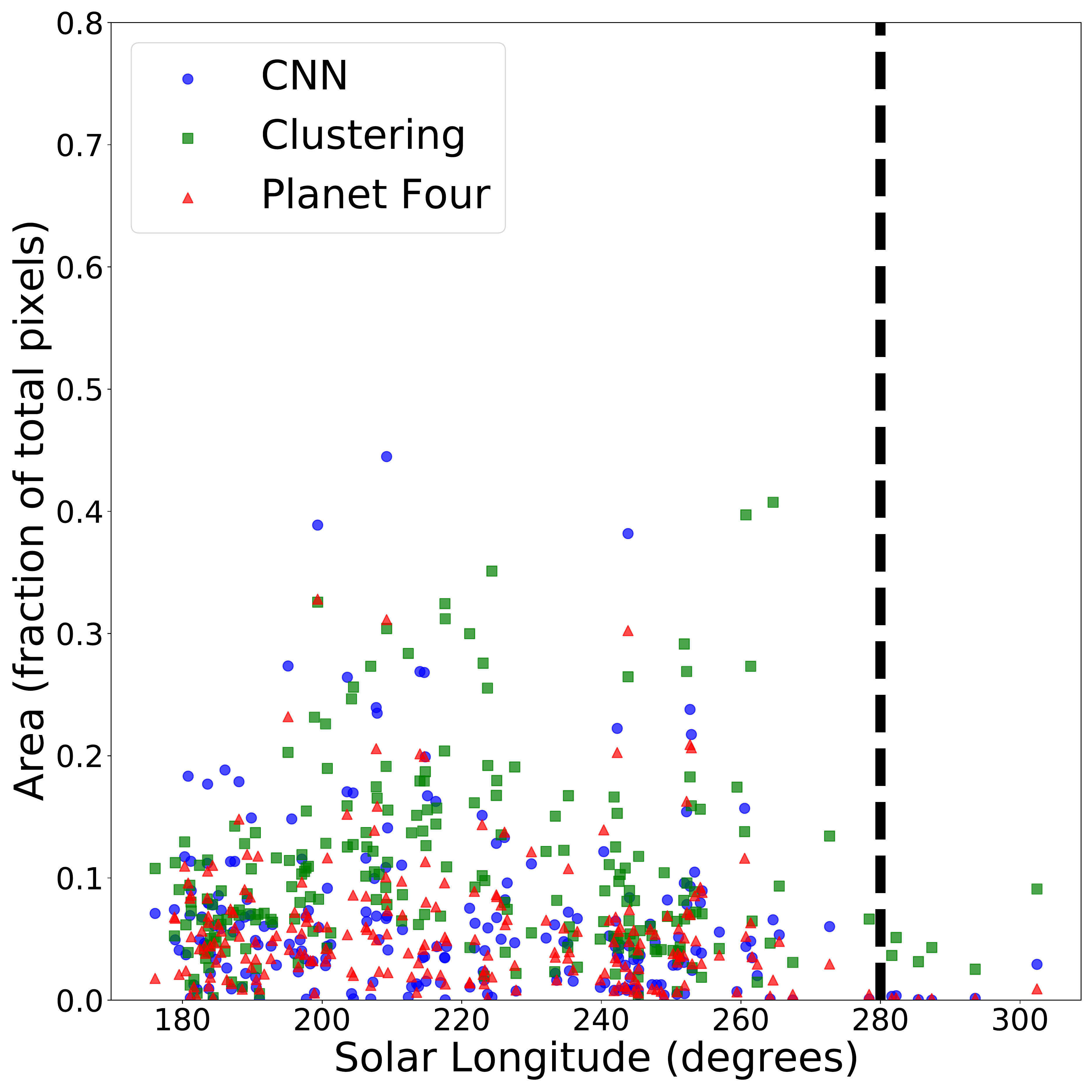}
\caption{{\bf Per-HiRISE-image:  Dice coefficient and area predicted as solar longitude varies.}  Each marker shows a result from one of the 221 HiRISE images. The thick dashed black line shows the solar longitude at which the ice-free transition occurs. The left-hand plot indicates that Dice coefficient tends to be low for all images at a high solar longitude. The right-hand plot shows the fraction of the number of pixels  in each image predicted as  CO$_2$ jet seasonal deposits. That this is lower for higher solar longitudes suggests that poorer Dice coefficient may be attributable to both algorithm and Planet Four volunteers finding it harder to identify  CO$_2$ jet seasonal deposits in nearly ice free images that consequently have fewer actual CO$_2$ jets. Both plots indicate that the CNN performs better than ISODATA clustering, with the right-hand plot suggesting ISODATA clustering over-predicts the area relative to the Planet Four volunteers.}
\label{fig:Res5}
\end{figure}

The HiRISE images each year were acquired during the Martian spring and early summer. At some point in this interval, the ice sheet sublimates fully away. It is possible that for images acquired when the ice sheet is close to fully sublimating that it is harder for Planet Four volunteers to identify  CO$_2$ jet seasonal deposits  than for the algorithms. To try to understand this, Figure~\ref{fig:Res5} (left) shows how Dice coefficient and total area predicted per HiRISE image varies with solar longitude for each of the 221 HiRISE images, from cross-validation results. There is a statistical negative correlation between Dice coefficient and solar longitude (Pearson correlation -0.22 for CNN and -0.31 for ISODATA clustering, with p values 0.0009 and 0.000001 respectively). In comparison any correlation with latitude is very low (Pearson correlation 0.08 for CNN and -0.001 for ISODATA clustering, with p values 0.25 and 0.98 respectively). Figure~\ref{fig:Res5} shows, for example, that no image with a solar longitude above 280 (relating to Mars south polar summer solstice) has a Dice Coefficient above 0.2.   This suggests the algorithms agree less with the P4 catalog at high solar longitudes. Figure~\ref{fig:Res5} (right) shows how the total area predicted per HiRISE Image (normalised by the total number of pixels in each image) varies with solar longitude. This data indicates that the total fraction predicted per image is lower at higher solar longitude, which would be consistent with Planet Four volunteers and algorithms finding labeling harder due to the near absence of ice at high longitudes.  However, it should be noted that because there are fewer fans/blotches labeled in the P4 catalogs for high solar longitudes, due to a lot less ice and hence CO$_2$ jets. There are also shadows in such images, which make it harder for human volunteers to determine markings.

\section{Supervised Learning for Binary Classification of Tiles}
\label{method-ml-classify}

For this secondary investigation, we designed a simpler supervised deep CNN for binary tile classification. The objective of this was to identify Planet Four tiles that were `empty', i.e.,~which did not contain any pixels that represent CO$_2$ jet seasonal deposits. For this purpose, instead of aiming to segment features, the CNN's input was an image tile of the same size as seen by human labelers, and its output was a binary prediction that either the tile contained no features at all, or otherwise. Although an alternative method might involve applying the semantic segmenter to each tile, this would mean that an an adhoc post-processing algorithm would need to be designed and optimised to account for small numbers of false positives in empty tiles. The advantage of designing a classifier is that it learns what the best prediction is for a tile in its entirety. % The clustering algorithm has not been tailored to address this problem so no results will be presented from that method.

\subsection{Design}%
\label{sec:design}
The training design for the CNN was very similar to the semantic segmenter. We used Leave-one-region-out cross-validation, and hence trained 28 models that were validated on all data from a single left-out region.  Data augmentation during training was used, including images rescaled to all 3 map scales, random horizontal and vertical flips, and 90 degree rotations. Such augmentation artificially increases the size of the training set; it is routinely carried out when training CNNs for computer vision tasks, as it helps combat overfitting. This data augmentation was not required for the Planet Four system as explained in Section \ref{sec:p4intro}.

Unlike our semantic segmentation approach, we made use of a pretrained model and transfer learning using the fine-tuning method~\citep{Kornblith.18}. We started with a ResNet-50~\citep{He.15a} pretrained on ImageNet~\citep{ILSVRC15} (available within tensorflow), and following typical practice for such a task, replaced its head with global average pooling and 2-class softmax layers~\citep{Goodfellow-et-al-2016,Kornblith.18}. The resulting network was fine tuned using stochastic gradient descent and cross-entropy loss, with a learning rate of 0.001 and momentum of 0.9, for ten epochs.  We trained three independently initialised networks, and for inference we averaged the classifier confidences produced for the positive class prior to calculating metrics.

%\subsubsection{Additional Metric for Binary Tile Classification}~\label{S:AUC}

For our binary tile classifiers, we additionally report results using the well known AUC (`Area Under the Curve') metric for binary classification~\citep{Ling_AUC}.  This metric can be used for binary classifiers that provide confidence values for each class for a given sample. AUC assesses how well the classifier performs on a validation dataset for each possible value of a decision threshold applied to the confidence value for the positive class. Ideally, all threshold values would provide a recall of 1.0 and a specificity of 1.0 (i.e.~no false negatives and no false positives). In this case, a plot of recall versus  sensitivity for each threshold value will have all data points at the coordinate $(1,1)$, which defines a shape with an area of 1.0.  In reality, as the threshold changes, recall will increase as specificity increases, and a changing threshold will define a curve starting at $(1,0)$ and ending at $(0,1)$, with an area less than 1. The AUC metric is a calculation of this area for a specific classifier applied to a specific validation dataset.

\subsection{Results}

\begin{figure}
\centering
\includegraphics[width=12cm]{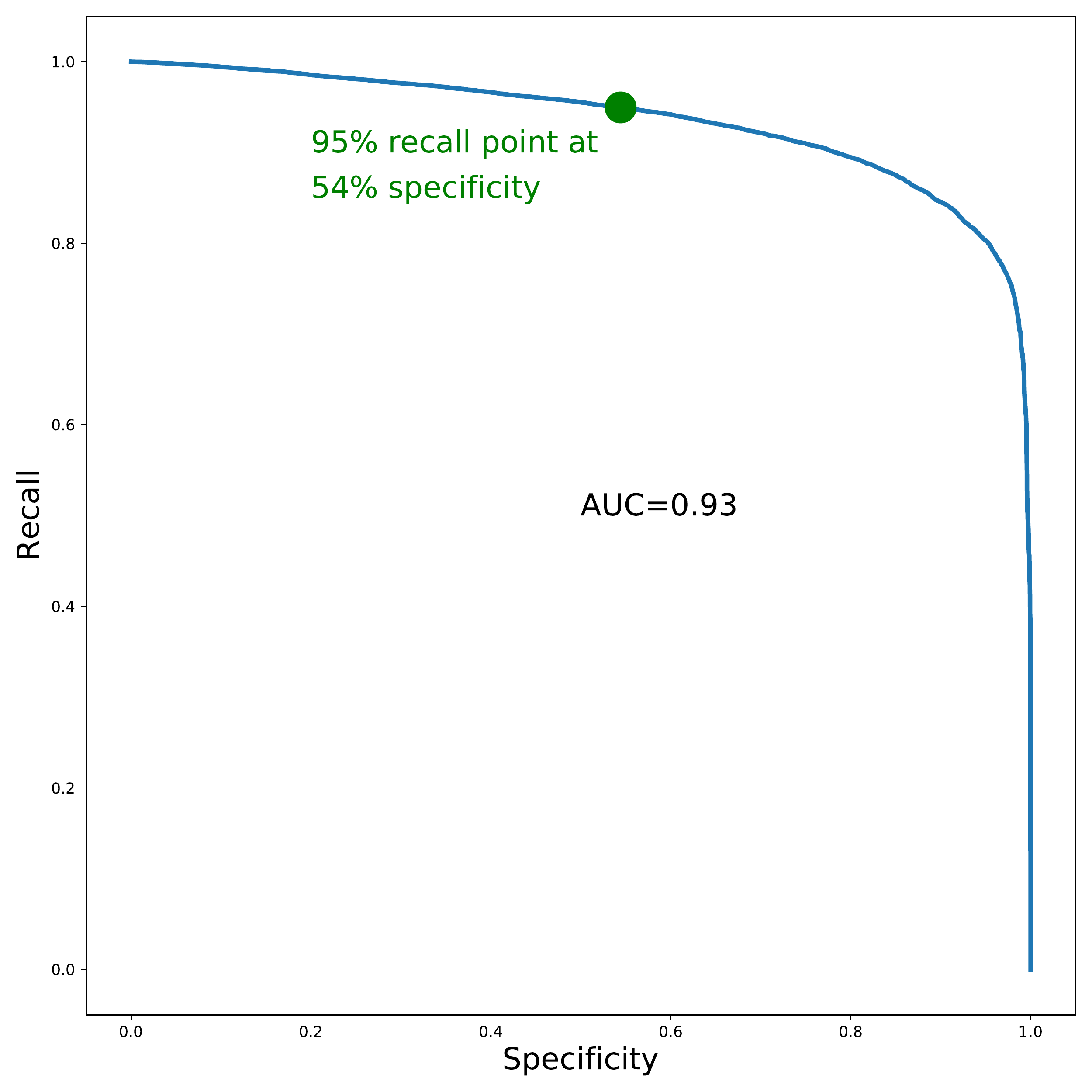}
\caption{{\bf Tradeoff between recall and specificity for binary empty image detection.} The AUC of our binary classification CNN is 0.93. The green circular marker indicates the operating point at which Recall=0.95. At this point, Specificity=0.54,~i.e. 54\%.}
\label{fig:AUC}
\end{figure}

In total, 42,904 tiles were classified in our leave-one-region-out cross-validation scheme for empty tile detection. The CNN we trained provides confidence values for each class as outputs, and hence we can evaluate it using AUC (see Section~\ref{method-ml-classify}); the AUC aggregated across all tiles was $0.93$. The relevant curve of recall vs specificity is shown in Figure~\ref{fig:AUC}.

We determined the binary classification threshold that would achieve 95\% recall (sensitivity) in our cross-validation data, i.e.~this target meant that 95\% of tiles with P4 catalog markings were correctly predicted by the classifier as having  CO$_2$ jet seasonal deposits. This required a decision threshold of $0.24$ to be applied to the classifier output's confidence for the  CO$_2$ jet seasonal deposits  class. At this value, the corresponding specificity was 0.54 (which means 54\% of tiles with no markings were correctly classified as having no  CO$_2$ jet seasonal deposits) the balanced accuracy was 0.75 (this is equivalent to the average of recall and specificity), and the precision was 0.87 (which means that 87\% of the tiles predicted to have  CO$_2$ jet seasonal deposits  by the classifier actually did have P4 catalog markings).  All these values can be confirmed by calculations on the Confusion Matrix (CM) at this decision point, which here we write in tabular form in Table~\ref{tab:CM}.  If our trained classifier was used in practice, and a higher recall or a higher specificity is required, the decision threshold can be changed accordingly.

\begin{table}[]
    \centering
    \begin{tabular}{|c|c|c|}
    \hline
         & Predicted No  CO$_2$    & Predicted  CO$_2$  \\
       &   jet seasonal deposits   & jet seasonal deposits \\
         \hline
   Planet Four catalog No Markings      &5388 &4513\\
   \hline
   Planet Four catalog Markings      &1661 &31342\\
   \hline
    \end{tabular}
    \caption{{\bf Confusion Matrix for binary tile classifier at Recall=0.95.} This data aggregates all tiles following our leave-one-region out cross-validation procedure.}
    \label{tab:CM}
\end{table}

\begin{figure}
\centering
\includegraphics[width=\columnwidth]{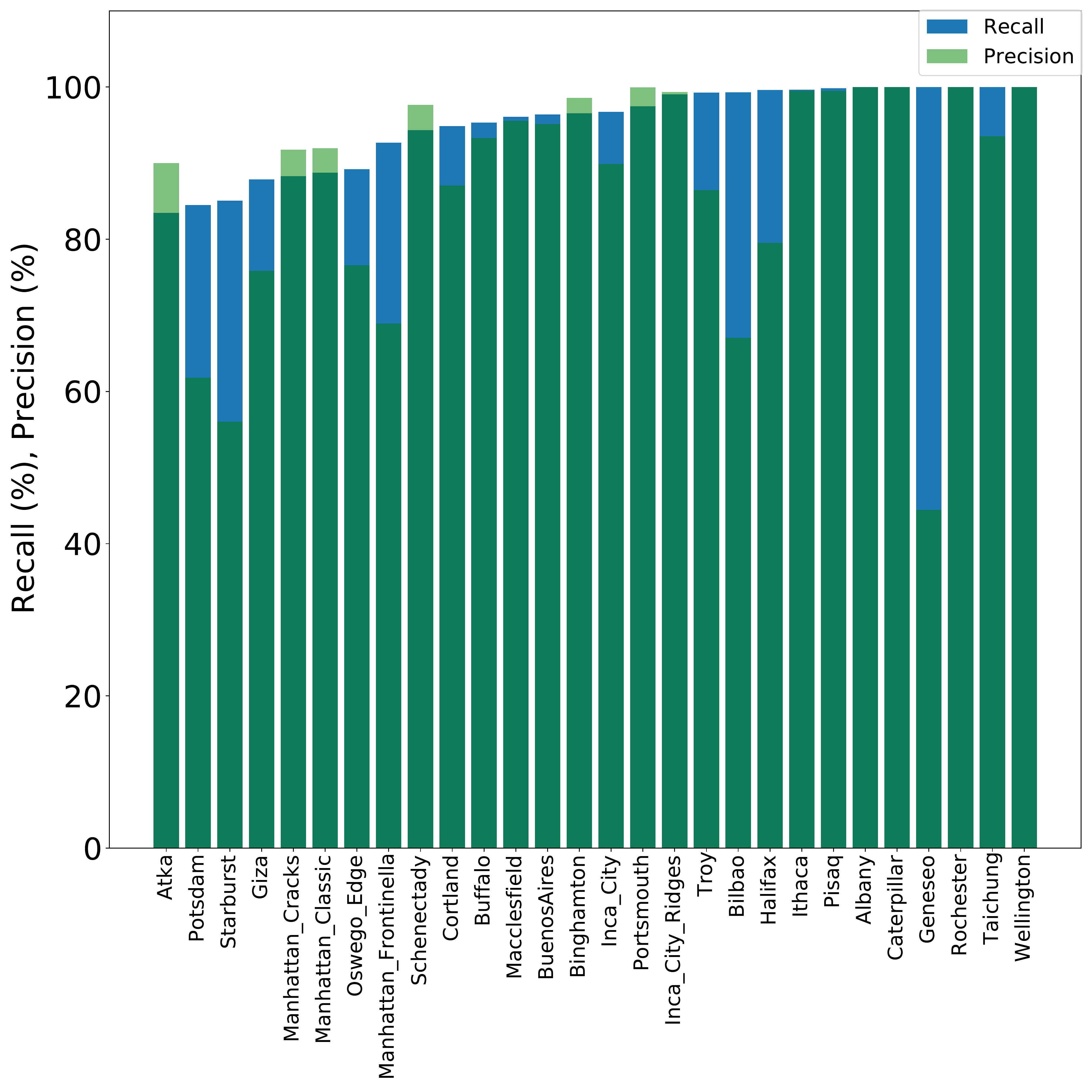}
\caption{{\bf Empty image detection by region.} Per-region recall and precision from leave-one-region-out cross validation, for a decision threshold of 0.24 (the value for which the overall recall is 0.95).}
\label{fig:classifier}
\end{figure}

To illustrate per-region performance, Figure~\ref{fig:classifier} shows the recall and precision results for a decision threshold of 0.24, when running our trained tile binary classifiers in leave-one-region-out cross-validation on tiles as seen by human labelers. The minimum recall is 0.84. However, the precision varies considerably by region, which suggests a need for more targeted filtering, such as by using different thresholds for different regions.

\clearpage

%%%%%%%%%%%%%%%%%%%%%%%%%%%%%%%%%%%%%%%%%%
\section{Discussion and Conclusions}\label{S:4}

\subsection{Using our methods to optimize human labeling}

Currently, all Planet Four tiles are reviewed by human volunteers, since its not known in advance which tiles have no  CO$_2$ jet seasonal deposits  on them. Therefore, a potential future use case for our binary tile classifier is to prioritise which tiles get shown to human labelers first, so as to minimize how often the Planet Four volunteers are  asked to annotate tiles that have no features. It would likely be desirable to have a high sensitivity while allowing a relatively high false positive rate.  For example, in our results, we achieved a specificity value of 0.54 at the chosen recall of 0.95. Hence, if only the model’s predictions determine what images to show to volunteers, this would mean our models could be used to `filter' approximately half of the tiles with no  CO$_2$ jet seasonal deposits  from the need for human inspection, while missing only 5\% of tiles that should have been labeled. The algorithm developed by \cite{Aye19} to generate the Planet Four fan and blotch catalog treats all the human volunteer assessments equally. The binary classifier and HRNet CNN results can be compared to the human volunteer classifications in order to identify the people who are particularly good at spotting that there are fans and blotches present in the HiRISE tiles. Then the classifications from these individuals could be weighted more heavily when the volunteer markings are combined together to produce the next edition of the Planet Four catalog. The latest version of the Planet Four project is now hosted on the Zooniverse's Project Builder Platform\footnote{\url{https://www.zooniverse.org/lab}}. The platform provides the functionality for easily combining the binary classifier with Planet Four human-generated classifications in real-time. The aggregated volunteer markings for Planet Four tiles under active review on the website could be compared to the binary classifier's label in order to identify which tiles need further volunteer review beyond the project's standard 30 independent reviews to more accurately map the positions and shapes of fans and blotches. We discuss this in further detail in Section \ref{sec:conclusions}.

\subsection{CNN Design Optimization}

In future work, it would be possible to improve on the performance of our HRNet by adding additional data augmentation (especially color, contrast and brightness augmentation), and possibly by enhancing low-contrast validation images. However, our intent in this paper was not to design the very best semantic segmentation model. Rather, we aimed to use a design that we knew from previous work performed well on binary segmentation in satellite imagery, and incorporated best practice machine learning principles used by default for mitigating the risk of overfitting. Hence, while we experimented briefly with various hyperparameters in the model, and adjusted until it was hard to improve performance, we did not seek to optimise them, nor to design the very best architecture.  

\subsection{Challenges for Supervised Learning with Ambiguous and Inconsistent Labels}\label{S4truth}

It needs to always be remembered that our HRNets are trained to try to replicate the Planet Four catalog's markings, and that our metrics are a comparison with the {\em agreement} with aggregated human labelers, rather than to an objective {\em ground-truth.} This was the main reason for not attempting to optimise the model to get better results than presented in this paper. Put another way the model was designed to predict what can be considered in a supervised learning context to be noisy and ambiguous labels~\citep{Algan.20}. The Planet Four catalog's fans and blotches have several known nonidealities from the supervised learning perspective, as follows. First, the choice of what image features to designate as a  CO$_2$ jet seasonal deposits  is subjective to an extent; the Planet Four catalog we used for training the HRNet is the result of aggregating labels from a varying number of humans, and required an algorithm to decide what the final catalog would contain in the event of human disagreement. Second, even when many humans agree a feature should be a  CO$_2$ jet seasonal deposit, the exact border around those deposits that different humans make is inconsistent. In particular, the border between CO$_2$ seasonal jets and ``background'' in HiRISE images is not sharp, i.e.~contrast in pixel values between the two categories can vary slowly spatially at boundaries. Moreover, humans were required to fit constrained ellipse and fan shapes to CO$_2$ seasonal jets, and this frequently introduces further inconsistency between which pixels are marked as belonging to which class. Third, some of the HiRISE image tiles were not completely annotated, and this would be a source of `confusion' to any supervised learning model.  The third aspect is one possible reason why the HRNet detected more total area than the P4 catalog. Compounding these challenges, our results for the center-overlap metric show that over 26\% of the centers of Planet Four catalog markings are missed by the CNN. Hence, our CNN model over-predicts total area relative to the Planet Four catalog, but under-predicts the total number of markings.

It should be noted that the training of the HRNet could easily be modified so its results are biased to prefer fewer false positives, and hence more closely match the total area of the P4 catalog overall, or to prefer fewer false negatives, and hence achieve a higher center-overlap. Either approach is likely, however, to reduce the overall Dice Coefficient, due to additional false negatives/positives created respectively. Moreover, our results also found substantial heterogeneity in log-area-ratio when this metric is broken down into per-region and per-image statistics, which suggests there is no one simple explanation for why the overall area predicted by the HRNet is larger than that of the P4 catalog.

From a machine learning perspective, all these reasons result in inconsistent pixel labels being provided during training for otherwise similar input features, which magnifies the degree of difficulty in learning to automatically identify CO$_2$ seasonal jets.

\subsection{Conclusions}
\label{sec:conclusions}
We examined computer-aided automated approaches to identifying wind-blown seasonal fans in the high-resolution imagery of the Martian south polar region. Leveraging the crowd-sourced catalog from the Planet Four project \citep{Aye19}, we successfully trained two types of deep CNN, a HRNet for semantic segmentation, and a binary per-image classifier. The HRNet was quite successful at identifying the darkened pixels that comprise seasonal fans and blotches within the HiRISE images, enabling accurate measurements of the changing surface area covered by seasonal fan and blotch material over time. The ISODATA clustering technique applied to the same task was less successful. It was able to identify some seasonal fans and blotches, but depending on the threshold selected, a visual inspection revealed the algorithm would often confuse some topographic features as being seasonal fans. 

A key aim of the Planet Four project is to measure the wind direction, inferred from the directions the fans are pointed in. The machine learning approaches we employed were not able to identify directions of the seasonal features. This remains an area where the crowd-sourced identifications excel compared to the automated algorithms. Future work will be focused on the automatic categorization of fans and blotches and measuring the directionality of these seasonal fans present in HiRISE images. Applying alternative deep CNNs and active learning techniques~\citep{10.1007/978-3-319-66179-7_46} with improved training and validation datasets may be able to succeed on this front. 

There is no objective ground-truth dataset of fans and blotches produced by carbon dioxide jets within the HiRISE images searched by the HRNet CNN. The Planet Four catalog has its own detection biases, and thus this work also serves as another validation of the \cite{Aye19} Planet Four seasonal fan and blotch catalog. The total number of pixels predicted by the HRNet CNN to belong to the markings class was 27$\%$ larger than the total number of pixels in fans/blotches in the Planet Four catalog, but this aspect varied considerably on a per-image basis. The majority of the discrepancies between the Planet Four catalog and the HRNet identifications occur within regions of the HiRISE frames where there are extremely high densities of seasonal fans and blotches (over 100 sources per Planet Four title or HiRISE subframe). In these cases, the volunteers marked different regions of the Planet Four tile or did not mark at all.  The true fraction of tiles with such high densities is likely small; 1.51$\%$ of the tiles have more than 100 fans and blotches recorded in the P4  catalog. Thus, the Planet Four catalog likely contains more than 70$\%$ of the seasonal CO$_2$ jet deposits visible in the majority of the HiRISE subframes.   
\subsection{Future Work}

Although Dice Coefficient, precision and recall are sensible measures for comparing outcomes from different machine learning algorithms, and the process of selecting the best algorithm, the underlying scientific question in the Planet Four project was to identify aspects such as wind direction and speed. For these questions, it is not necessarily important to measure the extent of agreement between the Planet Four catalog and machine learning. In the ideal case, predicted  CO$_2$ jet seasonal deposits  from  machine learning algorithms would be able to directly measure wind direction locally at all points in an image, using labeled fans as cues. Achieving this potentially requires new machine learning methods to be developed, as mentioned in the introduction. %Another use case for machine learning is to predict the total area of blotches and fans in an image region, such as a tile, with a higher area predicted potentially resulting in prioritisation for human labeling.

For the task of calculating the total area covered by fans or blotches, an alternative  to the semantic segmentation approach used here could be to use CNNs trained as object detectors. One challenge for a detector is the need to define thresholds during inference, in order to decide whether candidate detections should be retained as model outputs. For the dataset used in this paper, we would expect some difficulty in calibrating thresholds to work generally across regions. Another downside is that in some of the more densely populated images, detections would cover close to 100\% of the pixels in a large part of an image. Nevertheless, it would be interesting to attempt such an approach in future work.

There is also the potential for combining the machine learning developed here and crowd-sourced techniques, to maximize the effort of the volunteers performing the human review of high-resolution orbital imagery. For example, our machine learning algorithms detected some features missed in the hand-labeled dataset, and could provide an effective first pass over satellite images to determine whether any features are present, thereby potentially streamlining future delivery of images to citizen science projects, such as the Planet Four online platform.  A total of 29.6$\%$ (12,693) of the image tiles or HiRISE subframes that Planet Four volunteers spent time examining were found to be devoid of sources when the classifications were combined to create the Planet Four catalog~\citep{Aye19}. The binary classifier CNN could be deployed with the Planet Four project to significantly reduce the volunteer effort spent reviewing blank images with no seasonal fans or blotches. However, we note that removing all the blank images from the Planet Four project may also not be ideal, as~\cite{Bowyer2015} found that volunteer engagement decreased with the removal of blank images, with no animals present, from the Snapshot Serengeti project; see also~\cite{Jones20}.  Hence, as mentioned, it would be preferable to use the method not to remove all predicted blank images, but instead use it in a way that prioritises images that are predicted to contain markings, and/or reduce the number of images predicted to be blank.

As shown in Figure \ref{fig:P4Miss}, there is a small subset of Planet Four tiles where fans and blotches have been completely missed because there was no consensus amongst the human classifiers. This tends to be seen in images with a very large number of seasonal fans and blotches to individually mark. Volunteers do not mark all the features and either map a small number  of the visible fans and blotches or skip over marking completely. The Planet Four catalog is generated by finding where there is consensus amongst the volunteer markings in each tile. If the majority of human volunteers do not mark the same features visible in the title, then those fans and blotches will not be incorporated into the catalog. The binary classifier and/or the trained HRNet could be utilized to be identify which tiles are in need of additional human review. The output from the CNNs could be compared to the output from the aggregated human classifications. Tiles fans and blotches that come up as empty from the human-generated labels could be identified and made to receive additional human classifications on the Planet Four website until consensus is reached amongst the human reviewers. 

User weighting schemes has proven fruitful in a variety of crowdsourced astronomy projects with drawing tasks \citep[e.g.,][]{2012MNRAS.424.2442S, 2015ApJ...802..127J,2021MNRAS.501.4669E,2022arXiv220811760J}.  Instead of treating all the human-derived marks/drawing as equal, as is currently the case the \citep{Aye19} Planet Four catalog, the assessments from some volunteers are prioritized when the human-generated classifications are combined together to identify the features of interest. Clear criteria is needed to determine which human volunteers are excelling at the task and should be listened to more closely with increased user weights. The comparison or `gold standard' data is typically a small subset of the data reviewed in these citizen science projects. The more information available to assess the skill of the human volunteers, the bigger the expected impact a user weighting scheme will have on the aggregated results. The work presented here as potential applications in developing a user weighting scheme for Planet Four. The CNN results could be treated as ground truth to better identify those volunteers who are more adept at spotting the dark seasonal fans and blotches and outlining their shapes with the marking tools for future development of a user weighting scheme.

%\begin{itemize}
  %  \item Utility of the ML predicted masks for studying feature density per region, and the successive appearance of features across seasons .... i.e.  spatially projected masks, and  overlay masks of the same region from subsequent Mars years would reveal the changes in feature detection (eg. feature growth, appearance of new features). Hence the ML could be used for automated detection of new features not previously identified in a particular region.
  %  \item Challenges in describing feature geometry and thereby deriving wind vector parameters.
  %  \item ...a step towards autonomuous spacecraft and decision making to automatically target and take images without commands from humans...
%\end{itemize}

%%%%%%%%%%%%%%%%%%%%%%%%%%%%%%%%%%%%%%%%%%

\section*{Acknowledgements}

This publication uses data generated via the Zooniverse.org platform, development of which was supported by the Alfred P. Sloan Foundation. The Planet Four Catalog presented in this paper is the result of the efforts of the Planet Four volunteers, generously donating their time to the Planet Four project, and without whom this work would not have been possible. Their contributions are individually acknowledged at \url{http://www.planetfour.org/authors}. The authors also thank the anonymous referees for their feedback on the manuscript. 

This work is partially enabled by the NASA support for the Mars Reconnaissance Orbiter (MRO) High Resolution Imaging Science Experiment (HiRISE) team. This paper includes data collected by the MRO spacecraft and the HiRISE camera, and we gratefully acknowledge the entire MRO mission and HiRISE teams efforts in obtaining and providing the images used in this analysis. The authors thank HiRISE team members for multiple useful discussions and continuous support. The Mars Reconnaissance Orbiter mission is operated at the Jet Propulsion Laboratory, California Institute of Technology, under contracts with NASA. This work has also made use of NASA's Astrophysics Data System Bibliographic Services.

 KMA and GP were supported in part by National Aeronautics and Space Administration grants  NNX15AH36G and  80NSSC20K0748. MES is supported by the UK Science Technology Facilities Council (STFC) grant ST/V000691/1.

Data Access:  Our python code for training and validating our models, and producing Figures 6--14 in this paper, is available on Zenodo: 
\href{https://doi.org/10.5281/zenodo.4292195}{doi:10.5281/zenodo.4292195}.Information about accessing the Planet Four classification data, catalog, and associated HiRISE image information is discussed in \cite{Aye19}. 
%\software{Python 3.7~\citep{Python3},
%TensorFlow 2.1~\citep{tensorflow2015-whitepaper}, IDL~\citep{Didriksen}, ENVI~\citep{envitask}
%          }
\bibliographystyle{cas-model2-names}
\bibliography{mybibfile}

\begin{thebibliography}{54}
\expandafter\ifx\csname natexlab\endcsname\relax\def\natexlab#1{#1}\fi
\providecommand{\url}[1]{\texttt{#1}}
\providecommand{\href}[2]{#2}
\providecommand{\path}[1]{#1}
\providecommand{\DOIprefix}{doi:}
\providecommand{\ArXivprefix}{arXiv:}
\providecommand{\URLprefix}{URL: }
\providecommand{\Pubmedprefix}{pmid:}
\providecommand{\doi}[1]{\href{http://dx.doi.org/#1}{\path{#1}}}
\providecommand{\Pubmed}[1]{\href{pmid:#1}{\path{#1}}}
\providecommand{\bibinfo}[2]{#2}
\ifx\xfnm\relax \def\xfnm[#1]{\unskip,\space#1}\fi
%Type = Article
\bibitem[{Abdollahzadeh et~al.(2021)Abdollahzadeh, Sepehr and
  Rashki}]{Abdollahzadeh2021}
\bibinfo{author}{Abdollahzadeh, S.}, \bibinfo{author}{Sepehr, A.},
  \bibinfo{author}{Rashki, A.}, \bibinfo{year}{2021}.
\newblock \bibinfo{title}{Detecting degraded, prone and transition ecosystems
  by environmental thresholds and spectral functions}.
\newblock \bibinfo{journal}{Remote Sensing Applications: Society and
  Environment} \bibinfo{volume}{22}, \bibinfo{pages}{100503}.
%Type = Misc
\bibitem[{Algan and Ulusoy(2020)}]{Algan.20}
\bibinfo{author}{Algan, G.}, \bibinfo{author}{Ulusoy, I.},
  \bibinfo{year}{2020}.
\newblock \bibinfo{title}{Label noise types and their effects on deep
  learning}.
\newblock \href{http://arxiv.org/abs/2003.10471}{\tt arXiv:2003.10471}.
%Type = Article
\bibitem[{Aye et~al.(2019)Aye, Schwamb, Portyankina, Hansen, McMaster, Miller,
  Carstensen, Snyder, Parrish, Lynn, Mai, Miller, Simpson and Smith}]{Aye19}
\bibinfo{author}{Aye, K.M.}, \bibinfo{author}{Schwamb, M.E.},
  \bibinfo{author}{Portyankina, G.}, \bibinfo{author}{Hansen, C.J.},
  \bibinfo{author}{McMaster, A.}, \bibinfo{author}{Miller, G.R.M.},
  \bibinfo{author}{Carstensen, B.}, \bibinfo{author}{Snyder, C.},
  \bibinfo{author}{Parrish, M.}, \bibinfo{author}{Lynn, S.},
  \bibinfo{author}{Mai, C.}, \bibinfo{author}{Miller, D.},
  \bibinfo{author}{Simpson, R.J.}, \bibinfo{author}{Smith, A.M.},
  \bibinfo{year}{2019}.
\newblock \bibinfo{title}{Planet four: Probing springtime winds on mars by
  mapping the southern polar {CO2} jet deposits}.
\newblock \bibinfo{journal}{Icarus} \bibinfo{volume}{319},
  \bibinfo{pages}{558--598}.
\newblock \DOIprefix\doi{10.1016/j.icarus.2018.08.018}.
%Type = Book
\bibitem[{Ball and Hall(1965)}]{Ball}
\bibinfo{author}{Ball, G.H.}, \bibinfo{author}{Hall, D.J.},
  \bibinfo{year}{1965}.
\newblock \bibinfo{title}{Isodata: a method of data analysis and pattern
  classification}.
\newblock \bibinfo{publisher}{Stanford Research Institute}.
%Type = Article
\bibitem[{{Bickel} et~al.(2019){Bickel}, {Lanaras}, {Manconi}, {Loew} and
  {Mall}}]{Bickel19}
\bibinfo{author}{{Bickel}, V.T.}, \bibinfo{author}{{Lanaras}, C.},
  \bibinfo{author}{{Manconi}, A.}, \bibinfo{author}{{Loew}, S.},
  \bibinfo{author}{{Mall}, U.}, \bibinfo{year}{2019}.
\newblock \bibinfo{title}{Automated detection of lunar rockfalls using a
  convolutional neural network}.
\newblock \bibinfo{journal}{IEEE Transactions on Geoscience and Remote Sensing}
  \bibinfo{volume}{57}, \bibinfo{pages}{3501--3511}.
%Type = Article
\bibitem[{Bowyer et~al.(2015)Bowyer, Maidel, Lintott, Swanson and
  Miller}]{Bowyer2015}
\bibinfo{author}{Bowyer, A.}, \bibinfo{author}{Maidel, V.},
  \bibinfo{author}{Lintott, C.}, \bibinfo{author}{Swanson, A.},
  \bibinfo{author}{Miller, G.}, \bibinfo{year}{2015}.
\newblock \bibinfo{title}{This image intentionally left blank: Mundane images
  increase citizen science participation}.
\newblock \bibinfo{journal}{Conference on Human Computation and Crowdsourcing}
  \URLprefix \url{http://rgdoi.net/10.13140/RG.2.2.35844.53121},
  \DOIprefix\doi{10.13140/RG.2.2.35844.53121}.
%Type = Article
\bibitem[{Didriksen et~al.(1987)Didriksen, Lie and Conradi}]{Didriksen}
\bibinfo{author}{Didriksen, T.}, \bibinfo{author}{Lie, A.},
  \bibinfo{author}{Conradi, R.}, \bibinfo{year}{1987}.
\newblock \bibinfo{title}{Idl as a data description language for a programming
  environment database} \bibinfo{volume}{22}.
\newblock \URLprefix \url{https://doi.org/10.1145/39305.39312},
  \DOIprefix\doi{10.1145/39305.39312}.
%Type = Article
\bibitem[{{Eisner} et~al.(2021){Eisner}, {Barrag{\'a}n}, {Lintott}, {Aigrain},
  {Nicholson}, {Boyajian}, {Howell}, {Johnston}, {Lakeland}, {Miller},
  {McMaster}, {Parviainen}, {Safron}, {Schwamb}, {Trouille}, {Vaughan},
  {Zicher}, {Allen}, {Allen}, {Bouslog}, {Johnson}, {Simon}, {Wolfenbarger},
  {Baeten}, {Bundy} and {Hoffman}}]{2021MNRAS.501.4669E}
\bibinfo{author}{{Eisner}, N.L.}, \bibinfo{author}{{Barrag{\'a}n}, O.},
  \bibinfo{author}{{Lintott}, C.}, \bibinfo{author}{{Aigrain}, S.},
  \bibinfo{author}{{Nicholson}, B.}, \bibinfo{author}{{Boyajian}, T.S.},
  \bibinfo{author}{{Howell}, S.}, \bibinfo{author}{{Johnston}, C.},
  \bibinfo{author}{{Lakeland}, B.}, \bibinfo{author}{{Miller}, G.},
  \bibinfo{author}{{McMaster}, A.}, \bibinfo{author}{{Parviainen}, H.},
  \bibinfo{author}{{Safron}, E.J.}, \bibinfo{author}{{Schwamb}, M.E.},
  \bibinfo{author}{{Trouille}, L.}, \bibinfo{author}{{Vaughan}, S.},
  \bibinfo{author}{{Zicher}, N.}, \bibinfo{author}{{Allen}, C.},
  \bibinfo{author}{{Allen}, S.}, \bibinfo{author}{{Bouslog}, M.},
  \bibinfo{author}{{Johnson}, C.}, \bibinfo{author}{{Simon}, M.N.},
  \bibinfo{author}{{Wolfenbarger}, Z.}, \bibinfo{author}{{Baeten}, E.M.L.},
  \bibinfo{author}{{Bundy}, D.M.}, \bibinfo{author}{{Hoffman}, T.},
  \bibinfo{year}{2021}.
\newblock \bibinfo{title}{{Planet Hunters TESS II: findings from the first two
  years of TESS}}.
\newblock \bibinfo{journal}{\mnras} \bibinfo{volume}{501},
  \bibinfo{pages}{4669--4690}.
\newblock \DOIprefix\doi{10.1093/mnras/staa3739},
  \href{http://arxiv.org/abs/2011.13944}{\tt arXiv:2011.13944}.
%Type = Book
\bibitem[{Goodfellow et~al.(2016)Goodfellow, Bengio and
  Courville}]{Goodfellow-et-al-2016}
\bibinfo{author}{Goodfellow, I.}, \bibinfo{author}{Bengio, Y.},
  \bibinfo{author}{Courville, A.}, \bibinfo{year}{2016}.
\newblock \bibinfo{title}{Deep Learning}.
\newblock \bibinfo{publisher}{MIT Press}.
\newblock \bibinfo{note}{\url{http://www.deeplearningbook.org}}.
%Type = Article
\bibitem[{{Hansen} et~al.(2010){Hansen}, {Thomas}, {Portyankina}, {McEwen},
  {Becker}, {Byrne}, {Herkenhoff}, {Kieffer} and
  {Mellon}}]{2010Icar..205..283H}
\bibinfo{author}{{Hansen}, C.J.}, \bibinfo{author}{{Thomas}, N.},
  \bibinfo{author}{{Portyankina}, G.}, \bibinfo{author}{{McEwen}, A.},
  \bibinfo{author}{{Becker}, T.}, \bibinfo{author}{{Byrne}, S.},
  \bibinfo{author}{{Herkenhoff}, K.}, \bibinfo{author}{{Kieffer}, H.},
  \bibinfo{author}{{Mellon}, M.}, \bibinfo{year}{2010}.
\newblock \bibinfo{title}{{HiRISE observations of gas sublimation-driven
  activity in Mars' southern polar regions: I. Erosion of the surface}}.
\newblock \bibinfo{journal}{\icarus} \bibinfo{volume}{205},
  \bibinfo{pages}{283--295}.
\newblock \DOIprefix\doi{10.1016/j.icarus.2009.07.021}.
%Type = Inproceedings
\bibitem[{He et~al.(2017)He, Gkioxari, Dollár and Girshick}]{MaskRCNN}
\bibinfo{author}{He, K.}, \bibinfo{author}{Gkioxari, G.},
  \bibinfo{author}{Dollár, P.}, \bibinfo{author}{Girshick, R.},
  \bibinfo{year}{2017}.
\newblock \bibinfo{title}{{Mask R-CNN}}, in: \bibinfo{booktitle}{2017 IEEE
  International Conference on Computer Vision (ICCV)}, pp.
  \bibinfo{pages}{2980--2988}.
\newblock \DOIprefix\doi{10.1109/ICCV.2017.322}.
%Type = Techreport
\bibitem[{He et~al.(2015)He, Zhang, Ren and Sun}]{He.15a}
\bibinfo{author}{He, K.}, \bibinfo{author}{Zhang, X.}, \bibinfo{author}{Ren,
  S.}, \bibinfo{author}{Sun, J.}, \bibinfo{year}{2015}.
\newblock \bibinfo{title}{Deep residual learning for image recognition}.
\newblock \bibinfo{type}{Technical Report}. Microsoft Research.
\newblock \bibinfo{note}{Arxiv.1512.03385}.
%Type = Article
\bibitem[{Jadon(2020)}]{jadon2020survey}
\bibinfo{author}{Jadon, S.}, \bibinfo{year}{2020}.
\newblock \bibinfo{title}{A survey of loss functions for semantic
  segmentation}.
\newblock \bibinfo{journal}{arXiv preprint arXiv:2006.14822} .
%Type = Article
\bibitem[{{Johnson} et~al.(2015){Johnson}, {Seth}, {Dalcanton}, {Wallace},
  {Simpson}, {Lintott}, {Kapadia}, {Skillman}, {Caldwell}, {Fouesneau},
  {Weisz}, {Williams}, {Beerman}, {Gouliermis} and
  {Sarajedini}}]{2015ApJ...802..127J}
\bibinfo{author}{{Johnson}, L.C.}, \bibinfo{author}{{Seth}, A.C.},
  \bibinfo{author}{{Dalcanton}, J.J.}, \bibinfo{author}{{Wallace}, M.L.},
  \bibinfo{author}{{Simpson}, R.J.}, \bibinfo{author}{{Lintott}, C.J.},
  \bibinfo{author}{{Kapadia}, A.}, \bibinfo{author}{{Skillman}, E.D.},
  \bibinfo{author}{{Caldwell}, N.}, \bibinfo{author}{{Fouesneau}, M.},
  \bibinfo{author}{{Weisz}, D.R.}, \bibinfo{author}{{Williams}, B.F.},
  \bibinfo{author}{{Beerman}, L.C.}, \bibinfo{author}{{Gouliermis}, D.A.},
  \bibinfo{author}{{Sarajedini}, A.}, \bibinfo{year}{2015}.
\newblock \bibinfo{title}{{PHAT Stellar Cluster Survey. II. Andromeda Project
  Cluster Catalog}}.
\newblock \bibinfo{journal}{\apj} \bibinfo{volume}{802}, \bibinfo{pages}{127}.
\newblock \DOIprefix\doi{10.1088/0004-637X/802/2/127},
  \href{http://arxiv.org/abs/1501.04966}{\tt arXiv:1501.04966}.
%Type = Article
\bibitem[{{Johnson} et~al.(2022){Johnson}, {Wainer}, {TorresVillanueva},
  {Seth}, {Williams}, {Durbin}, {Dalcanton}, {Weisz}, {Bell}, {Guhathakurta},
  {Skillman} and {Smercina}}]{2022arXiv220811760J}
\bibinfo{author}{{Johnson}, L.C.}, \bibinfo{author}{{Wainer}, T.M.},
  \bibinfo{author}{{TorresVillanueva}, E.E.}, \bibinfo{author}{{Seth}, A.C.},
  \bibinfo{author}{{Williams}, B.F.}, \bibinfo{author}{{Durbin}, M.J.},
  \bibinfo{author}{{Dalcanton}, J.J.}, \bibinfo{author}{{Weisz}, D.R.},
  \bibinfo{author}{{Bell}, E.F.}, \bibinfo{author}{{Guhathakurta}, P.},
  \bibinfo{author}{{Skillman}, E.}, \bibinfo{author}{{Smercina}, A.},
  \bibinfo{year}{2022}.
\newblock \bibinfo{title}{{The Panchromatic Hubble Andromeda Treasury:
  Triangulum Extended Region (PHATTER). IV. Star Cluster Catalog}}.
\newblock \bibinfo{journal}{arXiv e-prints} ,
  \bibinfo{pages}{arXiv:2208.11760}\href{http://arxiv.org/abs/2208.11760}{\tt
  arXiv:2208.11760}.
%Type = Article
\bibitem[{Jones et~al.(2020)Jones, Arteta, Zisserman, Lempitsky, Lintott and
  Hart}]{Jones20}
\bibinfo{author}{Jones, F.M.}, \bibinfo{author}{Arteta, C.},
  \bibinfo{author}{Zisserman, A.}, \bibinfo{author}{Lempitsky, V.},
  \bibinfo{author}{Lintott, C.J.}, \bibinfo{author}{Hart, T.},
  \bibinfo{year}{2020}.
\newblock \bibinfo{title}{Processing citizen science- and machine-annotated
  time-lapse imagery for biologically meaningful metrics}.
\newblock \bibinfo{journal}{Scientific Data} \bibinfo{volume}{7},
  \bibinfo{pages}{102}.
%Type = Article
\bibitem[{Kaufmann and Hagermann(2017)}]{Kaufmann2017118}
\bibinfo{author}{Kaufmann, E.}, \bibinfo{author}{Hagermann, A.},
  \bibinfo{year}{2017}.
\newblock \bibinfo{title}{Experimental investigation of insolation-driven dust
  ejection from marsâ<u+0080><u+0099> \{CO2\} ice caps}.
\newblock \bibinfo{journal}{Icarus} \bibinfo{volume}{282}, \bibinfo{pages}{118
  -- 126}.
\newblock \DOIprefix\doi{http://dx.doi.org/10.1016/j.icarus.2016.09.039}.
%Type = Article
\bibitem[{{Kerner} et~al.(2019){Kerner}, {Wagstaff}, {Bue}, {Gray}, {Bell} and
  {Ben Amor}}]{Kerner.19}
\bibinfo{author}{{Kerner}, H.R.}, \bibinfo{author}{{Wagstaff}, K.L.},
  \bibinfo{author}{{Bue}, B.D.}, \bibinfo{author}{{Gray}, P.C.},
  \bibinfo{author}{{Bell}, James~F., I.}, \bibinfo{author}{{Ben Amor}, H.},
  \bibinfo{year}{2019}.
\newblock \bibinfo{title}{{Toward Generalized Change Detection on Planetary
  Surfaces With Convolutional Autoencoders and Transfer Learning}}.
\newblock \bibinfo{journal}{IEEE Journal of Selected Topics in Applied Earth
  Observations and Remote Sensing} \bibinfo{volume}{12},
  \bibinfo{pages}{3900--3918}.
\newblock \DOIprefix\doi{10.1109/JSTARS.2019.2936771}.
%Type = Inproceedings
\bibitem[{{Kieffer}(2000)}]{2000mpse.conf...93K}
\bibinfo{author}{{Kieffer}, H.H.}, \bibinfo{year}{2000}.
\newblock \bibinfo{title}{{Annual Punctuated CO2 Slab-Ice and Jets on Mars}},
  in: \bibinfo{booktitle}{Second International Conference on Mars Polar Science
  and Exploration}, p.~\bibinfo{pages}{93}.
%Type = Article
\bibitem[{{Kieffer}(2007)}]{2007JGRE..112.8005K}
\bibinfo{author}{{Kieffer}, H.H.}, \bibinfo{year}{2007}.
\newblock \bibinfo{title}{{Cold jets in the Martian polar caps}}.
\newblock \bibinfo{journal}{Journal of Geophysical Research (Planets)}
  \bibinfo{volume}{112}, \bibinfo{pages}{E08005}.
\newblock \DOIprefix\doi{10.1029/2006JE002816}.
%Type = Article
\bibitem[{{Kieffer} et~al.(2006){Kieffer}, {Christensen} and
  {Titus}}]{2006Natur.442..793K}
\bibinfo{author}{{Kieffer}, H.H.}, \bibinfo{author}{{Christensen}, P.R.},
  \bibinfo{author}{{Titus}, T.N.}, \bibinfo{year}{2006}.
\newblock \bibinfo{title}{{CO$_2$ jets formed by sublimation beneath
  translucent slab ice in Mars' seasonal south polar ice cap}}.
\newblock \bibinfo{journal}{\nat} \bibinfo{volume}{442},
  \bibinfo{pages}{793--796}.
\newblock \DOIprefix\doi{10.1038/nature04945}.
%Type = Article
\bibitem[{Kornblith et~al.(2018)Kornblith, Shlens and Le}]{Kornblith.18}
\bibinfo{author}{Kornblith, S.}, \bibinfo{author}{Shlens, J.},
  \bibinfo{author}{Le, Q.V.}, \bibinfo{year}{2018}.
\newblock \bibinfo{title}{Do better {I}mage{N}et models transfer better?}
\newblock \bibinfo{journal}{Arxiv:} \bibinfo{volume}{1805.08974}.
\newblock \URLprefix \url{http://arxiv.org/abs/1805.08974},
  \href{http://arxiv.org/abs/1805.08974}{\tt arXiv:1805.08974}.
%Type = Misc
\bibitem[{L3Harris(2020)}]{envitask}
\bibinfo{author}{L3Harris}, \bibinfo{year}{2020}.
\newblock \bibinfo{title}{Enviisodataclassificationtask}.
\newblock \URLprefix
  \url{https://www.harrisgeospatial.com/docs/enviisodataclassificationtask.html}.
%Type = Misc
\bibitem[{L3Harris(2022)}]{envitasklabel}
\bibinfo{author}{L3Harris}, \bibinfo{year}{2022}.
\newblock \bibinfo{title}{Label$\_$region}.
\newblock \URLprefix
  \url{https://www.l3harrisgeospatial.com/docs/label_region.html}.
%Type = Article
\bibitem[{Lee(2019)}]{Lee19}
\bibinfo{author}{Lee, C.}, \bibinfo{year}{2019}.
\newblock \bibinfo{title}{Automated crater detection on mars using deep
  learning}.
\newblock \bibinfo{journal}{Planetary and Space Science} \bibinfo{volume}{170},
  \bibinfo{pages}{16--28}.
\newblock \URLprefix
  \url{https://www.sciencedirect.com/science/article/pii/S0032063318303945},
  \DOIprefix\doi{https://doi.org/10.1016/j.pss.2019.03.008}.
%Type = Inproceedings
\bibitem[{Ling et~al.(2003)Ling, Huang and Zhang}]{Ling_AUC}
\bibinfo{author}{Ling, C.Z.}, \bibinfo{author}{Huang, J.},
  \bibinfo{author}{Zhang, H.}, \bibinfo{year}{2003}.
\newblock \bibinfo{title}{{AUC}: {A} better measure than accuracy in comparing
  learning algorithms}.
%Type = Inproceedings
\bibitem[{{Long} et~al.(2015){Long}, {Shelhamer} and {Darrell}}]{Long15}
\bibinfo{author}{{Long}, J.}, \bibinfo{author}{{Shelhamer}, E.},
  \bibinfo{author}{{Darrell}, T.}, \bibinfo{year}{2015}.
\newblock \bibinfo{title}{Fully convolutional networks for semantic
  segmentation}, in: \bibinfo{booktitle}{2015 IEEE Conference on Computer
  Vision and Pattern Recognition (CVPR)}, pp. \bibinfo{pages}{3431--3440}.
%Type = Article
\bibitem[{Mahboob and Genc(2019)}]{Mahboob2019}
\bibinfo{author}{Mahboob, M.}, \bibinfo{author}{Genc, B.},
  \bibinfo{year}{2019}.
\newblock \bibinfo{title}{Evaluation of isodata clustering algorithm for
  surface gold mining using satellite data}.
\newblock \bibinfo{journal}{2019 International Conference on Electrical,
  Communication, and Computer Engineering (ICECCE)} \bibinfo{volume}{July
  2019}.
%Type = Article
\bibitem[{McEwen et~al.(2007)McEwen, Eliason, Bergstrom, Bridges, Hansen,
  Delamere, Grant, Gulick, Herkenhoff, Keszthelyi, Kirk, Mellon, Squyres,
  Thomas and Weitz}]{McEwen07}
\bibinfo{author}{McEwen, A.S.}, \bibinfo{author}{Eliason, E.M.},
  \bibinfo{author}{Bergstrom, J.W.}, \bibinfo{author}{Bridges, N.T.},
  \bibinfo{author}{Hansen, C.J.}, \bibinfo{author}{Delamere, W.A.},
  \bibinfo{author}{Grant, J.A.}, \bibinfo{author}{Gulick, V.C.},
  \bibinfo{author}{Herkenhoff, K.E.}, \bibinfo{author}{Keszthelyi, L.},
  \bibinfo{author}{Kirk, R.L.}, \bibinfo{author}{Mellon, M.T.},
  \bibinfo{author}{Squyres, S.W.}, \bibinfo{author}{Thomas, N.},
  \bibinfo{author}{Weitz, C.M.}, \bibinfo{year}{2007}.
\newblock \bibinfo{title}{Mars reconnaissance orbiter's high resolution imaging
  science experiment (hirise)}.
\newblock \bibinfo{journal}{Journal of Geophysical Research: Planets}
  \bibinfo{volume}{112}.
\newblock \URLprefix
  \url{https://agupubs.onlinelibrary.wiley.com/doi/abs/10.1029/2005JE002605},
  \DOIprefix\doi{10.1029/2005JE002605},
  \href{http://arxiv.org/abs/https://agupubs.onlinelibrary.wiley.com/doi/pdf/10.1029/2005JE002605}{\tt
  arXiv:https://agupubs.onlinelibrary.wiley.com/doi/pdf/10.1029/2005JE002605}.
%Type = Article
\bibitem[{{McEwen} et~al.(2007){McEwen}, {Eliason}, {Bergstrom}, {Bridges},
  {Hansen}, {Delamere}, {Grant}, {Gulick}, {Herkenhoff}, {Keszthelyi}, {Kirk},
  {Mellon}, {Squyres}, {Thomas} and {Weitz}}]{2007JGRE..112.5S02M}
\bibinfo{author}{{McEwen}, A.S.}, \bibinfo{author}{{Eliason}, E.M.},
  \bibinfo{author}{{Bergstrom}, J.W.}, \bibinfo{author}{{Bridges}, N.T.},
  \bibinfo{author}{{Hansen}, C.J.}, \bibinfo{author}{{Delamere}, W.A.},
  \bibinfo{author}{{Grant}, J.A.}, \bibinfo{author}{{Gulick}, V.C.},
  \bibinfo{author}{{Herkenhoff}, K.E.}, \bibinfo{author}{{Keszthelyi}, L.},
  \bibinfo{author}{{Kirk}, R.o.L.}, \bibinfo{author}{{Mellon}, M.T.},
  \bibinfo{author}{{Squyres}, S.W.}, \bibinfo{author}{{Thomas}, N.},
  \bibinfo{author}{{Weitz}, C.M.}, \bibinfo{year}{2007}.
\newblock \bibinfo{title}{{Mars Reconnaissance Orbiter's High Resolution
  Imaging Science Experiment (HiRISE)}}.
\newblock \bibinfo{journal}{Journal of Geophysical Research (Planets)}
  \bibinfo{volume}{112}, \bibinfo{pages}{E05S02}.
\newblock \DOIprefix\doi{10.1029/2005JE002605}.
%Type = Article
\bibitem[{Minaee et~al.(2021)Minaee, Boykov, Porikli, Plaza, Kehtarnavaz and
  Terzopoulos}]{Minaee21}
\bibinfo{author}{Minaee, S.}, \bibinfo{author}{Boykov, Y.Y.},
  \bibinfo{author}{Porikli, F.}, \bibinfo{author}{Plaza, A.J.},
  \bibinfo{author}{Kehtarnavaz, N.}, \bibinfo{author}{Terzopoulos, D.},
  \bibinfo{year}{2021}.
\newblock \bibinfo{title}{Image segmentation using deep learning: A survey}.
\newblock \bibinfo{journal}{IEEE Transactions on Pattern Analysis and Machine
  Intelligence} ,
  \bibinfo{pages}{1--1}\DOIprefix\doi{10.1109/TPAMI.2021.3059968}.
%Type = Article
\bibitem[{{Pilorget} et~al.(2013){Pilorget}, {Edwards}, {Ehlmann}, {Forget} and
  {Millour}}]{2013JGRE..118.2520P}
\bibinfo{author}{{Pilorget}, C.}, \bibinfo{author}{{Edwards}, C.S.},
  \bibinfo{author}{{Ehlmann}, B.L.}, \bibinfo{author}{{Forget}, F.},
  \bibinfo{author}{{Millour}, E.}, \bibinfo{year}{2013}.
\newblock \bibinfo{title}{{Material ejection by the cold jets and temperature
  evolution of the south seasonal polar cap of Mars from THEMIS/CRISM
  observations and implications for surface properties}}.
\newblock \bibinfo{journal}{Journal of Geophysical Research (Planets)}
  \bibinfo{volume}{118}, \bibinfo{pages}{2520--2536}.
\newblock \DOIprefix\doi{10.1002/2013JE004513}.
%Type = Article
\bibitem[{Piqueux et~al.(2003)Piqueux, Byrne and Richardson}]{piqeux2003}
\bibinfo{author}{Piqueux, S.}, \bibinfo{author}{Byrne, S.},
  \bibinfo{author}{Richardson, M.}, \bibinfo{year}{2003}.
\newblock \bibinfo{title}{Sublimation of mars's southern seasonal co2 ice cap
  and the formation of spiders}.
\newblock \bibinfo{journal}{Journal of Geophysical Research: Planets}
  \bibinfo{volume}{108}.
%Type = Article
\bibitem[{{Piqueux} et~al.(2003){Piqueux}, {Byrne} and
  {Richardson}}]{2003JGRE..108.5084P}
\bibinfo{author}{{Piqueux}, S.}, \bibinfo{author}{{Byrne}, S.},
  \bibinfo{author}{{Richardson}, M.I.}, \bibinfo{year}{2003}.
\newblock \bibinfo{title}{{Sublimation of Mars's southern seasonal CO$_2$ ice
  cap and the formation of spiders}}.
\newblock \bibinfo{journal}{Journal of Geophysical Research (Planets)}
  \bibinfo{volume}{108}, \bibinfo{pages}{3--1}.
\newblock \DOIprefix\doi{10.1029/2002JE002007}.
%Type = Article
\bibitem[{{Piqueux} and {Christensen}(2008)}]{2008JGRE..113.6005P}
\bibinfo{author}{{Piqueux}, S.}, \bibinfo{author}{{Christensen}, P.R.},
  \bibinfo{year}{2008}.
\newblock \bibinfo{title}{{North and south subice gas flow and venting of the
  seasonal caps o f Mars: A major geomorphological agent}}.
\newblock \bibinfo{journal}{Journal of Geophysical Research (Planets)}
  \bibinfo{volume}{113}, \bibinfo{pages}{E06005}.
\newblock \DOIprefix\doi{10.1029/2007JE003009}.
%Type = Article
\bibitem[{Piqueux and Christensen(2008)}]{Piqueux08}
\bibinfo{author}{Piqueux, S.}, \bibinfo{author}{Christensen, P.R.},
  \bibinfo{year}{2008}.
\newblock \bibinfo{title}{North and south subice gas flow and venting of the
  seasonal caps of mars: A major geomorphological agent}.
\newblock \bibinfo{journal}{Journal of Geophysical Research: Planets}
  \bibinfo{volume}{113}.
\newblock \DOIprefix\doi{https://doi.org/10.1029/2007JE003009}.
%Type = Article
\bibitem[{{Portyankina} et~al.(2010){Portyankina}, {Markiewicz}, {Thomas},
  {Hansen} and {Milazzo}}]{2010Icar..205..311P}
\bibinfo{author}{{Portyankina}, G.}, \bibinfo{author}{{Markiewicz}, W.J.},
  \bibinfo{author}{{Thomas}, N.}, \bibinfo{author}{{Hansen}, C.J.},
  \bibinfo{author}{{Milazzo}, M.}, \bibinfo{year}{2010}.
\newblock \bibinfo{title}{{HiRISE observations of gas sublimation-driven
  activity in Mars' southern polar regions: III. Models of processes involving
  translucent ice}}.
\newblock \bibinfo{journal}{\icarus} \bibinfo{volume}{205},
  \bibinfo{pages}{311--320}.
\newblock \DOIprefix\doi{10.1016/j.icarus.2009.08.029}.
%Type = Inproceedings
\bibitem[{Rahman and Wang(2016)}]{Rahman16}
\bibinfo{author}{Rahman, M.A.}, \bibinfo{author}{Wang, Y.},
  \bibinfo{year}{2016}.
\newblock \bibinfo{title}{Optimizing intersection-over-union in deep neural
  networks for image segmentation}, in: \bibinfo{booktitle}{Proc. International
  Symposium on Visual Computing}, pp. \bibinfo{pages}{234--244}.
%Type = Inproceedings
\bibitem[{Ronneberger et~al.(2015)Ronneberger, Fischer and
  Brox}]{Ronneberger15}
\bibinfo{author}{Ronneberger, O.}, \bibinfo{author}{Fischer, P.},
  \bibinfo{author}{Brox, T.}, \bibinfo{year}{2015}.
\newblock \bibinfo{title}{U-net: Convolutional networks for biomedical image
  segmentation}, in: \bibinfo{editor}{Navab, N.}, \bibinfo{editor}{Hornegger,
  J.}, \bibinfo{editor}{Wells, W.M.}, \bibinfo{editor}{Frangi, A.F.} (Eds.),
  \bibinfo{booktitle}{Medical Image Computing and Computer-Assisted
  Intervention---MICCAI 2015}, \bibinfo{publisher}{Springer International
  Publishing}, \bibinfo{address}{Cham}. pp. \bibinfo{pages}{234--241}.
%Type = Article
\bibitem[{Russakovsky et~al.(2015)Russakovsky, Deng, Su, Krause, Satheesh, Ma,
  Huang, Karpathy, Khosla, Bernstein, Berg and Fei-Fei}]{ILSVRC15}
\bibinfo{author}{Russakovsky, O.}, \bibinfo{author}{Deng, J.},
  \bibinfo{author}{Su, H.}, \bibinfo{author}{Krause, J.},
  \bibinfo{author}{Satheesh, S.}, \bibinfo{author}{Ma, S.},
  \bibinfo{author}{Huang, Z.}, \bibinfo{author}{Karpathy, A.},
  \bibinfo{author}{Khosla, A.}, \bibinfo{author}{Bernstein, M.},
  \bibinfo{author}{Berg, A.C.}, \bibinfo{author}{Fei-Fei, L.},
  \bibinfo{year}{2015}.
\newblock \bibinfo{title}{{ImageNet Large Scale Visual Recognition Challenge}}.
\newblock \bibinfo{journal}{International Journal of Computer Vision (IJCV)}
  \bibinfo{volume}{115}, \bibinfo{pages}{211--252}.
\newblock \DOIprefix\doi{10.1007/s11263-015-0816-y}.
%Type = Inbook
\bibitem[{Sammut and Webb(2010)}]{LOO}
\bibinfo{editor}{Sammut, C.}, \bibinfo{editor}{Webb, G.I.} (Eds.),
  \bibinfo{year}{2010}.
\newblock \bibinfo{title}{Leave-One-Out Cross-Validation}.
  \bibinfo{publisher}{Springer US}, \bibinfo{address}{Boston, MA}.
\newblock pp. \bibinfo{pages}{600--601}.
%Type = Article
\bibitem[{Silburt et~al.(2019)Silburt, Ali-Dib, Zhu, Jackson, Valencia, Kissin,
  Tamayo and Menou}]{Silburt19}
\bibinfo{author}{Silburt, A.}, \bibinfo{author}{Ali-Dib, M.},
  \bibinfo{author}{Zhu, C.}, \bibinfo{author}{Jackson, A.},
  \bibinfo{author}{Valencia, D.}, \bibinfo{author}{Kissin, Y.},
  \bibinfo{author}{Tamayo, D.}, \bibinfo{author}{Menou, K.},
  \bibinfo{year}{2019}.
\newblock \bibinfo{title}{Lunar crater identification via deep learning}.
\newblock \bibinfo{journal}{Icarus} \bibinfo{volume}{317},
  \bibinfo{pages}{27--38}.
\newblock \URLprefix
  \url{https://www.sciencedirect.com/science/article/pii/S0019103518301386},
  \DOIprefix\doi{https://doi.org/10.1016/j.icarus.2018.06.022}.
%Type = Article
\bibitem[{{Simpson} et~al.(2012){Simpson}, {Povich}, {Kendrew}, {Lintott},
  {Bressert}, {Arvidsson}, {Cyganowski}, {Maddison}, {Schawinski}, {Sherman},
  {Smith} and {Wolf-Chase}}]{2012MNRAS.424.2442S}
\bibinfo{author}{{Simpson}, R.J.}, \bibinfo{author}{{Povich}, M.S.},
  \bibinfo{author}{{Kendrew}, S.}, \bibinfo{author}{{Lintott}, C.J.},
  \bibinfo{author}{{Bressert}, E.}, \bibinfo{author}{{Arvidsson}, K.},
  \bibinfo{author}{{Cyganowski}, C.}, \bibinfo{author}{{Maddison}, S.},
  \bibinfo{author}{{Schawinski}, K.}, \bibinfo{author}{{Sherman}, R.},
  \bibinfo{author}{{Smith}, A.M.}, \bibinfo{author}{{Wolf-Chase}, G.},
  \bibinfo{year}{2012}.
\newblock \bibinfo{title}{{The Milky Way Project First Data Release: a bubblier
  Galactic disc}}.
\newblock \bibinfo{journal}{\mnras} \bibinfo{volume}{424},
  \bibinfo{pages}{2442--2460}.
\newblock \DOIprefix\doi{10.1111/j.1365-2966.2012.20770.x},
  \href{http://arxiv.org/abs/1201.6357}{\tt arXiv:1201.6357}.
%Type = Article
\bibitem[{Sprinks et~al.(2019)Sprinks, Houghton, Bamford and
  Morley}]{Sprinks19}
\bibinfo{author}{Sprinks, J.}, \bibinfo{author}{Houghton, R.},
  \bibinfo{author}{Bamford, S.}, \bibinfo{author}{Morley, J.G.},
  \bibinfo{year}{2019}.
\newblock \bibinfo{title}{Planet four: Craters—optimizing task workflow to
  improve volunteer engagement and crater counting performance}.
\newblock \bibinfo{journal}{Meteoritics \& Planetary Science}
  \bibinfo{volume}{54}, \bibinfo{pages}{1325--1346}.
\newblock \URLprefix
  \url{https://onlinelibrary.wiley.com/doi/abs/10.1111/maps.13277},
  \DOIprefix\doi{https://doi.org/10.1111/maps.13277}.
%Type = Article
\bibitem[{Stroppiana et~al.(2019)Stroppiana, Villa, Sona, Ronchetti, Candiani,
  Pepe, Busetto, Migliazzi and Boschetti}]{Stroppiana2019}
\bibinfo{author}{Stroppiana, D.}, \bibinfo{author}{Villa, P.},
  \bibinfo{author}{Sona, G.}, \bibinfo{author}{Ronchetti, G.},
  \bibinfo{author}{Candiani, G.}, \bibinfo{author}{Pepe, M.},
  \bibinfo{author}{Busetto, L.}, \bibinfo{author}{Migliazzi, M.},
  \bibinfo{author}{Boschetti, M.}, \bibinfo{year}{2019}.
\newblock \bibinfo{title}{Early season weed mapping in rice crops using
  multi-spectral uav data}.
\newblock \bibinfo{journal}{International Journal of Remote Sensing}
  \bibinfo{volume}{39}, \bibinfo{pages}{5432--5452}.
%Type = Article
\bibitem[{{Thomas} et~al.(2010){Thomas}, {Hansen}, {Portyankina} and
  {Russell}}]{2010Icar..205..296T}
\bibinfo{author}{{Thomas}, N.}, \bibinfo{author}{{Hansen}, C.J.},
  \bibinfo{author}{{Portyankina}, G.}, \bibinfo{author}{{Russell}, P.S.},
  \bibinfo{year}{2010}.
\newblock \bibinfo{title}{{HiRISE observations of gas sublimation-driven
  activity in Mars southern polar regions: II. Surficial deposits and their
  origins}}.
\newblock \bibinfo{journal}{\icarus} \bibinfo{volume}{205},
  \bibinfo{pages}{296--310}.
\newblock \DOIprefix\doi{10.1016/j.icarus.2009.05.030}.
%Type = Misc
\bibitem[{Wagstaff et~al.(2021)Wagstaff, Lu, Dunkel, Grimes, Zhao, Cai, Cole,
  Doran, Francis, Lee and Mandrake}]{wagstaff2021mars}
\bibinfo{author}{Wagstaff, K.}, \bibinfo{author}{Lu, S.},
  \bibinfo{author}{Dunkel, E.}, \bibinfo{author}{Grimes, K.},
  \bibinfo{author}{Zhao, B.}, \bibinfo{author}{Cai, J.}, \bibinfo{author}{Cole,
  S.B.}, \bibinfo{author}{Doran, G.}, \bibinfo{author}{Francis, R.},
  \bibinfo{author}{Lee, J.}, \bibinfo{author}{Mandrake, L.},
  \bibinfo{year}{2021}.
\newblock \bibinfo{title}{Mars image content classification: Three years of
  nasa deployment and recent advances}.
\newblock \href{http://arxiv.org/abs/2102.05011}{\tt arXiv:2102.05011}.
%Type = Inproceedings
\bibitem[{Wagstaff et~al.(2018)Wagstaff, Lu, Stanboli, Grimes, Gowda and
  Padams}]{Wagstaff18}
\bibinfo{author}{Wagstaff, K.}, \bibinfo{author}{Lu, Y.},
  \bibinfo{author}{Stanboli, A.}, \bibinfo{author}{Grimes, K.},
  \bibinfo{author}{Gowda, T.}, \bibinfo{author}{Padams, J.},
  \bibinfo{year}{2018}.
\newblock \bibinfo{title}{Deep mars: Cnn classification of mars imagery for the
  pds imaging atlas}.
\newblock \URLprefix
  \url{https://www.aaai.org/ocs/index.php/AAAI/AAAI18/paper/view/16040/16400}.
%Type = Article
\bibitem[{Walmsley et~al.(2019)Walmsley, Smith, Lintott, Gal, Bamford,
  Dickinson, Fortson, Kruk, Masters, Scarlata, Simmons, Smethurst and
  Wright}]{Walmsley19}
\bibinfo{author}{Walmsley, M.}, \bibinfo{author}{Smith, L.},
  \bibinfo{author}{Lintott, C.}, \bibinfo{author}{Gal, Y.},
  \bibinfo{author}{Bamford, S.}, \bibinfo{author}{Dickinson, H.},
  \bibinfo{author}{Fortson, L.}, \bibinfo{author}{Kruk, S.},
  \bibinfo{author}{Masters, K.}, \bibinfo{author}{Scarlata, C.},
  \bibinfo{author}{Simmons, B.}, \bibinfo{author}{Smethurst, R.},
  \bibinfo{author}{Wright, D.}, \bibinfo{year}{2019}.
\newblock \bibinfo{title}{{Galaxy Zoo: probabilistic morphology through
  Bayesian CNNs and active learning}}.
\newblock \bibinfo{journal}{Monthly Notices of the Royal Astronomical Society}
  \bibinfo{volume}{491}, \bibinfo{pages}{1554--1574}.
\newblock \URLprefix \url{https://doi.org/10.1093/mnras/stz2816},
  \DOIprefix\doi{10.1093/mnras/stz2816},
  \href{http://arxiv.org/abs/https://academic.oup.com/mnras/article-pdf/491/2/1554/31144873/stz2816.pdf}{\tt
  arXiv:https://academic.oup.com/mnras/article-pdf/491/2/1554/31144873/stz2816.pdf}.
%Type = Article
\bibitem[{{Wang} et~al.(2020){Wang}, {Sun}, {Cheng}, {Jiang}, {Deng}, {Zhao},
  {Liu}, {Mu}, {Tan}, {Wang}, {Liu} and {Xiao}}]{Wang.20}
\bibinfo{author}{{Wang}, J.}, \bibinfo{author}{{Sun}, K.},
  \bibinfo{author}{{Cheng}, T.}, \bibinfo{author}{{Jiang}, B.},
  \bibinfo{author}{{Deng}, C.}, \bibinfo{author}{{Zhao}, Y.},
  \bibinfo{author}{{Liu}, D.}, \bibinfo{author}{{Mu}, Y.},
  \bibinfo{author}{{Tan}, M.}, \bibinfo{author}{{Wang}, X.},
  \bibinfo{author}{{Liu}, W.}, \bibinfo{author}{{Xiao}, B.},
  \bibinfo{year}{2020}.
\newblock \bibinfo{title}{Deep high-resolution representation learning for
  visual recognition}.
\newblock \bibinfo{journal}{IEEE Transactions on Pattern Analysis and Machine
  Intelligence} \bibinfo{volume}{Accepted}, \bibinfo{pages}{1--1}.
\newblock \DOIprefix\doi{10.1109/TPAMI.2020.2983686}.
%Type = Article
\bibitem[{Wu et~al.(2018)Wu, Wong, Rudnick, Shabala, Alger, Banfield, Ong,
  White, Garon, Norris, Andernach, Tate, Lukic, Tang, Schawinski and
  Diakogiannis}]{Wu18}
\bibinfo{author}{Wu, C.}, \bibinfo{author}{Wong, O.I.},
  \bibinfo{author}{Rudnick, L.}, \bibinfo{author}{Shabala, S.S.},
  \bibinfo{author}{Alger, M.J.}, \bibinfo{author}{Banfield, J.K.},
  \bibinfo{author}{Ong, C.S.}, \bibinfo{author}{White, S.V.},
  \bibinfo{author}{Garon, A.F.}, \bibinfo{author}{Norris, R.P.},
  \bibinfo{author}{Andernach, H.}, \bibinfo{author}{Tate, J.},
  \bibinfo{author}{Lukic, V.}, \bibinfo{author}{Tang, H.},
  \bibinfo{author}{Schawinski, K.}, \bibinfo{author}{Diakogiannis, F.I.},
  \bibinfo{year}{2018}.
\newblock \bibinfo{title}{{Radio Galaxy Zoo: Claran – a deep learning
  classifier for radio morphologies}}.
\newblock \bibinfo{journal}{Monthly Notices of the Royal Astronomical Society}
  \bibinfo{volume}{482}, \bibinfo{pages}{1211--1230}.
\newblock \URLprefix \url{https://doi.org/10.1093/mnras/sty2646},
  \DOIprefix\doi{10.1093/mnras/sty2646},
  \href{http://arxiv.org/abs/https://academic.oup.com/mnras/article-pdf/482/1/1211/26205089/sty2646.pdf}{\tt
  arXiv:https://academic.oup.com/mnras/article-pdf/482/1/1211/26205089/sty2646.pdf}.
%Type = Article
\bibitem[{Yang et~al.(2020)Yang, Zhao, Bruzzone, Benediktsson, Liang, Liu,
  Zeng, Guan, Li and Ouyang}]{Yang20}
\bibinfo{author}{Yang, C.}, \bibinfo{author}{Zhao, H.},
  \bibinfo{author}{Bruzzone, L.}, \bibinfo{author}{Benediktsson, J.A.},
  \bibinfo{author}{Liang, Y.}, \bibinfo{author}{Liu, B.},
  \bibinfo{author}{Zeng, X.}, \bibinfo{author}{Guan, R.}, \bibinfo{author}{Li,
  C.}, \bibinfo{author}{Ouyang, Z.}, \bibinfo{year}{2020}.
\newblock \bibinfo{title}{Lunar impact crater identification and age estimation
  with {C}hang’{E} data by deep and transfer learning}.
\newblock \bibinfo{journal}{Nature Communications} \bibinfo{volume}{11},
  \bibinfo{pages}{6358}.
\newblock \DOIprefix\doi{https://doi.org/10.1038/s41467-020-20215-y}.
%Type = Inproceedings
\bibitem[{Yang et~al.(2017)Yang, Zhang, Chen, Zhang and
  Chen}]{10.1007/978-3-319-66179-7_46}
\bibinfo{author}{Yang, L.}, \bibinfo{author}{Zhang, Y.}, \bibinfo{author}{Chen,
  J.}, \bibinfo{author}{Zhang, S.}, \bibinfo{author}{Chen, D.Z.},
  \bibinfo{year}{2017}.
\newblock \bibinfo{title}{Suggestive annotation: A deep active learning
  framework for biomedical image segmentation}, in:
  \bibinfo{editor}{Descoteaux, M.}, \bibinfo{editor}{Maier-Hein, L.},
  \bibinfo{editor}{Franz, A.}, \bibinfo{editor}{Jannin, P.},
  \bibinfo{editor}{Collins, D.L.}, \bibinfo{editor}{Duchesne, S.} (Eds.),
  \bibinfo{booktitle}{Medical Image Computing and Computer Assisted
  Intervention---MICCAI 2017}, \bibinfo{publisher}{Springer International
  Publishing}, \bibinfo{address}{Cham}. pp. \bibinfo{pages}{399--407}.
%Type = Article
\bibitem[{Yuan et~al.(2021)Yuan, Shi and Gu}]{Yuan21}
\bibinfo{author}{Yuan, X.}, \bibinfo{author}{Shi, J.}, \bibinfo{author}{Gu,
  L.}, \bibinfo{year}{2021}.
\newblock \bibinfo{title}{A review of deep learning methods for semantic
  segmentation of remote sensing imagery}.
\newblock \bibinfo{journal}{Expert Systems with Applications}
  \bibinfo{volume}{169}, \bibinfo{pages}{114417}.
\newblock \URLprefix
  \url{https://www.sciencedirect.com/science/article/pii/S0957417420310836},
  \DOIprefix\doi{https://doi.org/10.1016/j.eswa.2020.114417}.

\end{thebibliography}

\end{document}